\documentclass{aa}
\usepackage{geometry}
\usepackage{graphicx}
\usepackage{txfonts}
\usepackage{todonotes}
\usepackage{times}
\usepackage{amssymb}
\usepackage{amsmath, bm}
\usepackage{bmpsize}
\usepackage{epstopdf}
\usepackage{dblfloatfix} 
\graphicspath{{./}}
\usepackage[normalem]{ulem}
\usepackage[T1]{fontenc}
\usepackage{makecell}
\usepackage{multirow}
\usepackage{mwe}
\usepackage{mathtools}
\usepackage{comment}
\usepackage{natbib,twoopt}
\usepackage[colorlinks=true,allcolors=blue, breaklinks=true]{hyperref} 

\bibpunct{(}{)}{;}{a}{}{,}             
\makeatletter
  \newcommandtwoopt{\citeads}[3][][]{\href{http://adsabs.harvard.edu/abs/#3}%
    {\def\hyper@linkstart##1##2{}%
     \let\hyper@linkend\@empty\citealp[#1][#2]{#3}}}
  \newcommandtwoopt{\citepads}[3][][]{\href{http://adsabs.harvard.edu/abs/#3}%
    {\def\hyper@linkstart##1##2{}%
     \let\hyper@linkend\@empty\citep[#1][#2]{#3}}}
  \newcommandtwoopt{\citetads}[3][][]{\href{http://adsabs.harvard.edu/abs/#3}%
    {\def\hyper@linkstart##1##2{}%
     \let\hyper@linkend\@empty\citet[#1][#2]{#3}}}
  \newcommandtwoopt{\citeyearads}[3][][]%
    {\href{http://adsabs.harvard.edu/abs/#3}
    {\def\hyper@linkstart##1##2{}%
     \let\hyper@linkend\@empty\citeyear[#1][#2]{#3}}}
\makeatother

\newcommand{\hbympc}{~h ~\mathrm{Mpc^{-1}}}
\newcommand{\mpcbyh}{~h^{-1} ~\mathrm{Mpc}}
\newcommand{\HI}{\rm H{\sc i}}
\newcommand{\HII}{\rm H{\sc ii }}

\newcommand{\TB}{\delta T_{\mathrm{b}}}
\newcommand{\AVTB}{\overline{\delta T}_{\mathrm{b}}}
\newcommand{\MSUN}{{\mathrm{M}}_{\odot}}

\newcommand{\XHI}{x_{\mathrm{HI}}}
\newcommand{\AVXHI}{\overline{x}_{\mathrm{HI}}}
\newcommand{\AVXHII}{\overline{x}_{\mathrm{HII}}}
\newcommand{\AVXHIIMAX}{\overline{x}_{\mathrm{HII,max}}}

\newcommand{\XHII}{x_{\mathrm{HII}}}
\newcommand{\TS}{T_{\mathrm{S}}}
\newcommand{\TK}{T_{\mathrm{K}}}
\newcommand{\AVTK}{\overline{T}_{\mathrm{K}}}
\newcommand{\TCMB}{T_{\gamma}}
\newcommand{\TEFF}{T_{\gamma, {\mathrm{eff}}}}
\newcommand{\lya}{\mathrm{{Ly{\alpha}}}}
\newcommand{\OmegaB}{\Omega_{\mathrm{B}}}
\newcommand{\Omegam}{\Omega_{\mathrm{m}}}

\newcommand{\DTB}{\Delta^2}
\newcommand{\DTBLOFARMEAN}{\Delta^2_\mathrm{21}} \newcommand{\DTBLOFARSIGMA}{\Delta^2_\mathrm{21,err}}
\newcommand{\DTBLOFARSIGMASQ}{\Delta^4_\mathrm{21,err}}
\newcommand{\DTBMODELMEAN}{\Delta^2_\mathrm{m}}
\newcommand{\DTBMODELSIGMA}{\Delta^2_\mathrm{m,err}}
\newcommand{\DTBMODELSIGMASQ}{\Delta^4_\mathrm{m,err}}
\newcommand{\DTBANSMEAN}{\Delta^2_\mathrm{21-an}} \newcommand{\DTBANSSIGMA}{\Delta^2_\mathrm{21-an,err}}
\newcommand{\MMIN}{M_{\mathrm{min}}}
\newcommand{\MMINX}{M_{\mathrm{min,X}}}
\newcommand{\FX}{f_{X}}
\newcommand{\FHEAT}{f_{\mathrm{heat}}}
\newcommand{\RPEAK}{R^{\mathrm{heat}}_{\mathrm{peak}}}
\newcommand{\RFWHM}{\Delta R^{\mathrm{heat}}_{\mathrm{FWHM}}}

\begin{document}


\title{Constraints on the state of the IGM at $z\sim 8-10$ using redshifted 21-cm observations with LOFAR}
\titlerunning{Inferring the EoR IGM using LOFAR multi-redshift results}
\author{R. Ghara\thanks{\email{ghara.raghunath@gmail.com}}\inst{1},
S. Zaroubi\inst{2,3},
B. Ciardi\inst{4},
G. Mellema\inst{5},
S. K. Giri\inst{6,7},
F. G. Mertens\inst{2,8},
M. Mevius\inst{9},
L. V. E. Koopmans\inst{2},
I. T. Iliev\inst{10},
A. Acharya\inst{4},
S. A. Brackenhoff\inst{2},
E. Ceccotti\inst{2, 11},
K. Chege\inst{2},
I. Georgiev\inst{3,5},
S. Ghosh\inst{2},
I. Hothi\inst{8,12},
C. H\"{o}fer\inst{2},
Q. Ma\inst{13},
S. Munshi\inst{2},
A. R. Offringa\inst{9},
A. K. Shaw\inst{14},
V. N. Pandey\inst{9},
S. Yatawatta\inst{9}
\and
M. Choudhury\inst{15}}

\institute{Department of Physical Sciences, Indian Institute of Science Education and Research Kolkata, Mohanpur, WB 741 246, India
\and 
Kapteyn Astronomical Institute, University of Groningen, PO Box 800, 9700AV Groningen, The Netherlands
\and
ARCO (Astrophysics Research Center), Department of Natural Sciences, The Open University of Israel, 1 University Road, PO Box 808, Ra'anana 4353701, Israel
\and
Max-Planck Institute for Astrophysics, Karl-Schwarzschild-Stra{\ss}e 1, 85748 Garching, Germany
\and
The Oskar Klein Centre, Department of Astronomy, Stockholm University, AlbaNova, SE-10691 Stockholm, Sweden
\and
Van Swinderen Institute for Particle Physics and Gravity, University of Groningen, Nijenborgh 4, 9747 AG Groningen, The Netherlands
\and
Nordita, KTH Royal Institute of Technology and Stockholm University, Hannes Alfvéns väg 12, SE-106 91 Stockholm, Sweden
\and
LUX, Observatoire de Paris, PSL Research University, CNRS, Sorbonne Université, F-75014 Paris, France
\and
Astron, PO Box 2, 7990 AA Dwingeloo, the Netherlands
\and
Astronomy Centre, Department of Physics and Astronomy, Pevensey II Building, University of Sussex, Brighton BN1 9QH, UK
\and 
INAF -- Istituto di Radioastronomia, Via P.~Gobetti 101, 40129 Bologna, Italy
\and
Laboratoire de Physique de l’ENS, ENS, Universit\'{e} PSL, CNRS, Sorbonne Universit\'{e}, Universit\'{e}e Paris Cit\'{e}, 75005 Paris, France
\and
School of Physics and Electronic Science, Guizhou Normal University, Guiyang 550001, PR China
\and
Department of Computer Science, University of Nevada, Las Vegas, Nevada 89154, USA
\and
Center for Fundamental Physics of the Universe, Department of Physics, Brown University, Providence 02914, RI, USA
}

\authorrunning{LOFAR-EoR}

\date{Received XXX; accepted YYY}

\abstract
{The power spectra of the redshifted 21-cm signal from the Epoch of Reionization (EoR) contain information about the ionization and thermal states of the intergalactic medium (IGM), and depend on the properties of the sources that existed during that period. Recently, \citet{lofar2024} has analysed 10 nights of LOFAR high-band data and estimated upper limits on the 21-cm power spectrum at redshifts 8.3, 9.1 and 10.1. Here we use these upper limit results to constrain the properties of the IGM at those redshifts. We focus on the properties of the ionized and heated regions where the temperature is larger than that of the Cosmic Microwave Background (CMB). We model the power spectrum of the 21-cm signal with the code {\sc grizzly}, and use a Bayesian inference framework to explore the source parameters for uniform priors on their ranges. The framework also provides information about the IGM properties in the form of derived parameters. We do not include constraints from other observables except some very conservative limits on the maximum ionization fraction at those redshifts, estimated from the CMB Thomson scattering optical depth. In a model which includes a radio background in excess of the CMB, the 95 (68) per cent credible intervals of {\it disfavoured models} at redshift 9.1 for the chosen priors correspond to IGM states with averaged ionization and heated fraction below 0.46 ($\lesssim 0.05$), an average gas temperature below 44 K (4 K), and a characteristic size of the heated region $\lesssim 14 ~\mpcbyh$ ($\lesssim 3 ~\mpcbyh$). The 68 per cent credible interval suggests an excess radio background which is more than 100 per cent of the CMB at 1.42 GHz, while the 95 per cent credible interval of the radio background efficiency parameter spans the entire prior range. The behaviour of the credible intervals is similar at all redshifts. The models disfavoured by the LOFAR upper limits are extreme ones, as they are mainly driven by rare and large ionized or heated regions. We find that the inclusion of upper limits from other radio interferometric observations in the Bayesian analysis significantly increases the number of disfavoured EoR models, enhancing the disfavoured credible intervals of the IGM parameters, especially those related to the average gas temperature and size distribution of the heated regions. While our constraints are not yet very strong, upcoming more stringent results from 21-cm observations together with the detection of many high-$z$ galaxies with for example the James Webb Space Telescope will strengthen our understanding of this crucial phase of our Universe.}


\maketitle
\section{Introduction}
\label{sec_intro}

The formation of the first stars and galaxies transformed the Universe from a cold and neutral phase to a hot and highly ionized one. The era when ultra-violet (UV) radiation from the earliest sources ionized the neutral hydrogen (\HI) in the intergalactic medium (IGM) is known as the Epoch of Reionization (EoR). While concrete evidence for the existence of this phase transition exists, its details remain unknown. Several observational probes, such as the CMB Thomson scattering optical depth \citep[e.g.,][]{2020A&A...641A...6P}, high-redshift ($z$) $\lya$ emitters \citep[e.g.,][]{Hu10, Morales_2021, 2023MNRAS.526.1657T, Witten_2024}, the Gunn-Peterson trough and the $\lya$ damping wings in high-$z$ quasar spectra \citep[e.g.,][]{Fan06b, Becker_2015, 2018Natur.553..473B}, have provided limited but useful insight about the evolution of the average neutral fraction.  
    
Currently, the redshifted 21-cm signal from neutral hydrogen is the most promising observational probe of the spatial distribution and temporal evolution of \HI\ in the IGM \citep[see e.g.,][]{madau1997, Furlanetto2006, 2013ASSL..396...45Z}. This signal carries a large amount of information on the physical state of the IGM, as well as on the properties of the first UV and X-ray emitting sources. Indeed, as it is a line emission, the 21-cm signal permits us to study physical processes at different stages of the EoR. 

To measure the redshifted 21-cm signal, several radio observations have been set up or proposed, and can be divided into two categories.  The first type uses a single radio antenna and attempts to measure the time evolution of the sky-averaged brightness temperature ($\AVTB$) of the 21-cm signal.  Into this category fall instruments such as Experiment to Detect the Global EoR Signature \citep[EDGES;][]{2010Natur.468..796B}, EDGES2 \citep{monsalve2017, EDGES2018}, the Shaped Antenna measurement of the background RAdio Spectrum \citep[SARAS;][]{2015ApJ...801..138P}, SARAS2 \citep{singh2017}, Broadband Instrument for Global HydrOgen ReioNisation Signal \citep[BigHorns;][]{2015PASA...32....4S}, the Sonda Cosmologica de las Islas para la Deteccion de Hidrogeno Neutro  \citep[SciHi;][]{2014ApJ...782L...9V}, the Large Aperture Experiment to detect the Dark Ages \citep[LEDA;][]{price2018}, Radio Experiment for the Analysis of Cosmic Hydrogen \citep[REACH;][]{2022NatAs...6..984D}, Probing Radio Intensity at High-Z from Marion \citep[Prizm;][]{philip2018} and Mapper of the IGM Spin Temperature \citep[MIST;][]{Monsalve_2024}.

The second type uses large radio interferometers for measuring the spatial fluctuations of the redshifted \HI\ signal in a statistical sense, mainly through its power spectrum. To this category belong ongoing or completed observations such as the Low Frequency Array \citep[LOFAR\footnote{\url{http://www.lofar.org/}};][]{vanHaarlem2013LOFAR:ARray, 2017ApJ...838...65P}, the Giant Metrewave Radio Telescope \citep[GMRT;][]{paciga13}, the Precision Array for Probing the Epoch of Reionization \citep[PAPER\footnote{\url{http://eor.berkeley.edu/}};][]{parsons13, 2019ApJ...883..133K}, the Murchison Widefield Array \citep[MWA\footnote{\url{http://www.LOFARtelescope.org/}};][]{tingay13}, the Hydrogen Epoch of Reionization Array \citep[HERA\footnote{\url{https://reionization.org/}};][]{2017PASP..129d5001D} and the New Extension in Nançay Upgrading LOFAR \citep[NenuFAR\footnote{\url{https://nenufar.obs-nancay.fr/en/homepage-en/}};][]{munshi2024}. Around the start of the next decade the Square Kilometre Array (SKA)\footnote{\url{http://www.skatelescope.org/}} will become available, and its transformative sensitivity should also enable tomographic imaging of the signal \citep{mellema13,2015aska.confE..10M, Koopmans_2015, ghara16}.

These observations face a range of challenges. Among them is the mitigation of the galactic and extra-galactic foregrounds, which are 4-5 orders of magnitude brighter than the expected signal \citep[see, e.g.,][]{ghosh12}. However, the smooth frequency dependence of the foreground signal allows its subtraction \citep{2009MNRAS.397.1138H, 2015MNRAS.447.1973B, 2016MNRAS.458.2928C, 2018MNRAS.478.3640M, 2021MNRAS.500.2264H}, avoidance \citep{2010ApJ...724..526D, 2014PhRvD..90b3019L} or suppression \citep{kanan2007MNRAS.382..809D, ghara15c}. At the same time, an accurate calibration during the data analysis process is required to minimize the artefacts from bright foreground sources  \citep{2016MNRAS.461.3135B, 2017ApJ...838...65P}. Recent improvements in foregrounds mitigation algorithms \citep[e.g.,][]{2014PhRvD..90b3019L, 2018MNRAS.478.3640M, 2024MNRAS.527.3517M, 2024MNRAS.527.7835A, 2024arXiv241216853G}, calibration methods \citep[see, e.g.,][]{yatawatta20112011, 2019ApJ...884..105K, 2020ApJ...888...70K, 2020Mevius, Gan2022, 2023A&A...669A..20G}, mitigation of radio frequency interference \citep[see e.g.,][]{offringa2019, Wilensky_2019, 2025arXiv250321728M} and ionospheric effects \citep[e.g.,][]{2016RaSc...51..927M, 2021A&A...652A..37E}, as well as improved sky-model building \citep[see e.g.,][]{Patil2016, Ewall2017} and inclusion of polarisation effects \citep[e.g.,][]{2010MNRAS.409.1647J, Spinelli2018} in the data analysis step, are showing promising results in reducing the effects of systematics, and thus in helping to recover the targetted 21-cm signal.

So far only one detection of $\AVTB$ with a minimum at $z\approx 17$  \citep{EDGES2018} has been claimed based on an EDGES2 low-band observation. This, though, remains disputed \citep[e.g.\ in][] {2018Natur.564E..32H, 2018ApJ...858L..10D, 2019ApJ...880...26S, 2019ApJ...874..153B}, and requires confirmation from other similar observations. A recent observation in the 55-85 MHz band with SARAS3 is inconsistent with the results from \citet{EDGES2018}, ruling out the deep $\AVTB$ profile around $z\sim17$ with 95\% confidence \citep{singh2022}. Other global 21-cm signal experiments, such as LEDA \citep{2016MNRAS.461.2847B}, EDGES high-band \citep{monsalve2017, 2019ApJ...875...67M}, and SARAS2 \citep{singh2017}, have only provided upper limits on the signal amplitude.  Results from SARAS and EDGES, though, have started ruling out EoR scenarios and putting constraints on the early source properties, dark matter models, and strength of radio background \citep[e.g.,][]{2018Natur.555...71B, 2018PhRvL.121a1101F, 2018Natur.557..684M, 2019JCAP...04..051N, 2019MNRAS.487.3560C, 2020MNRAS.492..634G, 2020MNRAS.496.1445C, 2022JCAP...03..055G, 2023JApA...44...10B}.

This paper focuses on the results of radio interferometric observations of the 21-cm signal power spectrum $\DTB$, which contains much more information than the global signal. At the current time, such observations have only produced upper limits. A $2\sigma$ upper limit of $(248)^2~ \rm mK^2$ for $k=0.5 ~\hbympc$ at $z=8.6$ from the GMRT was the first of its kind \citep{paciga13}. Later PAPER\footnote{The initial strong upper limits from the PAPER collaboration \citep{2016ApJ...833..102B} was later revised to a weaker upper bound after taking into account signal loss in their analysis \citep{2018ApJ...868...26C, 2019ApJ...883..133K}.} \citep{parsons13, 2015ApJ...809...61A}, LOFAR  \citep{2017ApJ...838...65P, 2020MNRAS.493.1662M}, MWA \citep{2019ApJ...884....1B, 2020MNRAS.493.4711T} and HERA \citep{Abdurashidova_2023} produced additional upper limits. To date, $\DTB(k = 0.075 \hbympc, ~z=9.1, \rm 141 ~hours) \approx (73)^2 ~{\rm mK}^2$, $\DTB(k = 0.34 \hbympc, ~z=7.9, 1128 ~\rm hours) \approx {(21.4)}^{2} ~{\rm mK}^2$ and $\DTB(k=0.14 \hbympc,~z=6.5, 110 \rm ~hours)\approx (43)^2 ~{\rm mK}^2$ are the best upper limits obtained from LOFAR \citep{2020MNRAS.493.1662M}, HERA \citep{Abdurashidova_2023} and MWA \citep{2020MNRAS.493.4711T} EoR observations, respectively. 
Recent observations with the LOFAR-Low Band Antenna array \citep{2019MNRAS.488.4271G}, the Owens Valley Radio Observatory Long Wavelength Array  \citep[OVRO-LWA;][]{2019AJ....158...84E} and the NenuFAR \citep{munshi2024} produced upper limits at higher redshifts. These data have been used to rule out not only some unusual reionization scenarios requiring e.g. atypical cooling mechanisms or the presence of a radio background in addition to the CMB, but also less exotic ones \citep[see e.g.,][]{2020MNRAS.493.4728G, 2020arXiv200603203G, 2020MNRAS.498.4178M, 2022ApJ...924...51A}.

\citet{lofar2024} have recently published new results from the LOFAR team in the redshift range 8.3 -- 10.1, with a best 2$\sigma$ upper limit of $\Delta^2(k = 0.076~\hbympc, z=9.1, 140~\mathrm{hours})\approx (54.3)^2 \rm ~mK^2$. This is approximately a factor of two improvement at the same $k-$ scale and redshift in comparison to the previous one from  \citet{2020MNRAS.493.1662M}, and thus is expected to rule out even more EoR scenarios than those already excluded in \citet{2020MNRAS.493.4728G}, \citet{2020arXiv200603203G}, and \citet{2020MNRAS.498.4178M}.  Additionally, the upper limits at the other two redshifts might help in ruling out even more EoR scenarios.

It should be noted that the 21-cm signal measurements do not directly probe astrophysical sources, but rather the ionization and thermal state of the IGM. However, presenting the 21-cm signal observables in terms of IGM-related parameters is challenging \citep[see e.g.,][]{2022MNRAS.514.2010M, 2024A&A...687A.252G, choudhury2024}. As the 21-cm signal simulation codes use the source parameters as input, previous studies mainly focused on constraining the astrophysical source properties using either Fisher matrices or Bayesian inference techniques \citep[e.g.,][]{2016MNRAS.458.2710E, Park2019InferringSignal, 2020MNRAS.495.4845C}. It should be realised that the inference on the astrophysical source properties is limited by the assumptions of the source model used in the 21-cm signal simulation codes. We therefore put less emphasis on the inference results for the source parameters, and focus on the IGM properties instead.

\begin{table*}
\centering
\caption[]{Recent LOFAR upper limit on $\DTB$ at different $k$-bins for $z=8.3$, 9.1 and 10.1. The different columns show $k-$bins, mean value $\DTBLOFARMEAN (k, z)$ and corresponding $1\sigma$ error $ \DTBLOFARSIGMA (k, z)$. These are achieved from 140 hours of the NCP field data (see \citealt{lofar2024} for details).} 
\small
\tabcolsep 6pt
\renewcommand\arraystretch{1.5}
   \begin{tabular}{|c c c|c c c|c c c|}
\hline
\multicolumn{3}{|c | }{$z=8.3$  } & \multicolumn{3}{c | }{$z=9.1$  } & \multicolumn{3}{c | }{$z=10.1$  } \\
\hline
$k$ ($\hbympc$) & $\DTBLOFARMEAN$ (mK$^2$) & $ \DTBLOFARSIGMA$ (mK$^2$) & $k$ ($\hbympc$) & $\DTBLOFARMEAN$ (mK$^2$) & $ \DTBLOFARSIGMA$ (mK$^2$) & $k$ ($\hbympc$) & $\DTBLOFARMEAN$ (mK$^2$) & $ \DTBLOFARSIGMA$ (mK$^2$)  \\
\hline
0.083 & $(47.4)^2$ & $(31.7)^2$  & 0.076 & $(39.7)^2$ & $(24.1)^2$  & 0.076  & $(50.7)^2$ & $(28.5)^2$  \\
0.106 & $(77.7)^2$ & $(31.3)^2$  & 0.101 & $(62.1)^2$ & $(26.3)^2$  & 0.101  & $(75.5)^2$ & $(36.3)^2$  \\
0.138 & $(110.1)^2$ & $(35.8)^2$  & 0.133 & $(96.8)^2$ & $(36.8)^2$  & 0.133  & $(112.4)^2$ & $(46.8)^2$  \\
0.184 & $(158.4)^2$ & $(46.6)^2$  & 0.181 & $(150.1)^2$ & $(53.8)^2$  & 0.181  & $(172.4)^2$ & $(70.0)^2$  \\
0.242 & $(234.8)^2$ & $(62.4)^2$  & 0.240 & $(217.3)^2$ & $(73.2)^2$  & 0.240  & $(243.4)^2$ & $(93.1)^2$  \\
0.323 & $(303.3)^2$ & $(67.4)^2$  & 0.319 & $(283.6)^2$ & $(80.9)^2$  & 0.323  & $(302.8)^2$ & $(98.3)^2$  \\
0.436 & $(338.1)^2$ & $(80.8)^2$  & 0.432 & $(320.3)^2$ & $(84.3)^2$ & 0.434  & $(351.2)^2$ & $(94.4)^2$  \\

\hline
\end{tabular}
\label{tab_obs}
\end{table*}

In this paper, we use the new LOFAR upper limits to investigate which EoR scenarios are disfavoured within the redshift range of 8.3--10.1. We achieve this by employing the Bayesian inference framework previously used in \citet{2021MNRAS.503.4551G} to interpret multi-redshift upper limits from MWA observations \citep{2020MNRAS.493.4711T}. The main difference between this version of the framework and the one used in \citet{2020MNRAS.493.4728G} is that we consider the presence of an excess radio background component in addition to the CMB, similar to what was done in \citet{2020MNRAS.498.4178M}.

The paper is structured as follows. In Sect. \ref{sec:methodology}, we describe the new LOFAR upper limits used in this study and introduce the basic methodology of our Bayesian interpretation framework.  We present our results in Sect. \ref{sec:results}, and finally conclude our study in Sect. \ref{sec:con}. We adopt the cosmological parameters $\Omegam=0.27$, $\Omega_\Lambda=0.73$, $\OmegaB=0.044$, $h=0.7$ \citep[Wilkinson ~Microwave ~Anisotropy ~Probem WMAP;][]{2013ApJS..208...19H}, which are the same as used in the $N$-body simulations employed here.


\section{Methodology}
\label{sec:methodology}
Here we briefly describe the new LOFAR upper limits on the 21-cm signal power spectrum and give an overview of the framework employed to constrain the IGM properties with them. 

\subsection{Observations}
\label{sec:obs}

The LOFAR-EoR Key Science Project team has analysed data for the North Celestial Pole (NCP) field from 14 nights of observations taken during LOFAR Cycles 0 to 3. Previously, \citet{2020MNRAS.493.1662M} analysed 12 nights (resulting in 141 hours) of NCP data in the frequency range 134--146 MHz, and presented upper limits on the 21-cm signal power spectrum at $z=9.1$. In the latest results, the team has added two more nights at 134--146 MHz ($z=[8.73, 9.6]$), and also analysed two additional frequency bands: 122--134 MHz ($z=[9.6, 11.64]$) and 146--158 MHz ($z=[ 8, 8.73]$), while only the best 10 nights (corresponding to 140 hours) of data for each redshift bin are finally combined. This has resulted in new upper limits on the 21-cm signal power spectrum at $z=8.3$, 9.1 and 10.1. 

The details of the observations and the analysis techniques can be found in \citet{lofar2024}. Here we summarize the final results in Table \ref{tab_obs}. It shows the mean power spectrum $\DTBLOFARMEAN (k, z)$ and associated 1$\sigma$ error $ \DTBLOFARSIGMA (k, z)$. The best result is at $z=9.1$, with a $2\sigma$ upper limit of $(54.3~\rm mK)^2$  at  $k = 0.075 \hbympc$, which is a factor of $\approx$2 lower than the one reported in \citet{2020MNRAS.493.1662M}. This is caused by improvements in the data analysis pipeline (see \citealt{lofar2024} for more details).


\subsection{Interpretation Framework}
\label{sec:method}
The framework employed to constrain the IGM properties using the upper limit results from \citet{lofar2024} relies on a large ($\sim 10^5$) database of reionization scenarios, as well as the associated IGM properties and 21-cm signal power spectra. These theoretical power spectra are then used in a Bayesian analysis in combination with the measured ones to find the exclusion likelihood for a reionization scenario. The same approach was used in \citet{2021MNRAS.503.4551G} to interpret upper limits from observations with MWA. We refer the reader to \citet{2020MNRAS.493.4728G, 2021MNRAS.503.4551G} for a detailed description of the framework. 

The large database of reionization scenarios and associated 21-cm signals is produced with {\sc grizzly} \citep{ghara15a, ghara18}, an independent implementation of the {\sc bears} algorithm  \citep{Thom08, Thom09, Thom11, 2018NewA...64....9K}. It requires cosmological density and velocity fields, together with the dark matter halo catalogues as input. These are typically obtained from a cosmological N-body simulation. Given a set of astrophysical source parameters, it then employs a one-dimensional radiative transfer scheme to generate ionization fraction, gas temperature and $\lya$ flux fields. The N-body simulation as well as the source model used in {\sc grizzly} are described in the following sub-sections.

\subsubsection{N-body simulation}
\label{sec:simul}

The N-body outputs used here are the same employed in \cite{2020MNRAS.493.4728G} and were obtained from the PRACE\footnote{Partnership for Advanced Computing in Europe: \url{http://www.prace-ri.eu/}} project PRACE4LOFAR. The dark matter-only simulation was run using the {\sc cubep$^3$m} \citep{Harnois12} code with particle mass resolution of  $4.05\times 10^7~\MSUN$. The simulation cube has a size of $500 \mpcbyh$, corresponding to a field-of-view of $4.3^{\circ}\times 4.3^{\circ}$ at $z\approx$ 9. This length is enough to cover the largest scale probed by the LOFAR observations, which corresponds to a $k$-scale of $0.05 \hbympc$, requiring a size of at least $\approx 250 \mpcbyh$.

The density and velocity fields were smoothed onto a  $300^3$ grid \citep[see e.g,][]{2019MNRAS.489.1590G, 2019JCAP...02..058G}. The lists of dark matter halos were obtained on the fly during the N-body simulation using a spherical overdensity halo finder \citep{Watson2013TheAges}. The halo lists consist of halos with more than 25 particles, which leads to a minimum dark matter halo mass $\approx 10^9~\MSUN$. 

\subsubsection{ {\sc grizzly} source model}
\label{sec:grizzly}
The next step is to employ {\sc grizzly} \citep{ghara15a, ghara18} to generate 3D cubes of the 21-cm signal at the probed redshifts. For this, we use different combinations of astrophysical parameters, which encode the source properties 
in terms of the emissivity of UV, X-ray, $\lya$ and radio photons, as each of these affects the spatial fluctuations of the signal \citep[see e.g.,][]{Ross2019, Eide2020, Reis2020Highredshift}. Our source model assumes that the stellar mass $M_\star$ in a dark matter halo with mass  $M_{\rm halo}$  is $M_\star = f_\star ~ \frac{\OmegaB}{\Omegam} ~ M_{\rm halo}$, where $f_\star$ is the star formation efficiency. In all our simulations we used $f_\star=$ 0.02, which is consistent with studies such as \cite{2015ApJ...799...32B} and \cite{2016MNRAS.460..417S}.  

The source parameters which we vary to create our set of simulations are as follows:

\begin{itemize}
    \item Ionization efficiency ($\zeta$): This parameter regulates the emission rate of ionizing photons escaping into the IGM from a halo\footnote{The amount of ionizing photons reaching the IGM depends on various quantities such as the star formation rate, the stellar initial mass function, the intrinsic ionizing photons emission rate, and the escape fraction. The parameter $\zeta$ combines the effects of all these.}. We assume it to be constant per unit stellar mass, or in other words, this emission rate scales linearly with halo mass. The rate is then given by $\dot N_i=\zeta\times 2.85\times 10^{45}  ~{\rm s^{-1}} ~\MSUN^{-1}$, where we vary ${\log}_{10}$($\zeta$) in the range [-3, 3]. 
    
    \item X-ray heating efficiency ($f_X$): This is the parameter that regulates the emission rate of X-ray photons from a halo according to $\dot N_X = f_X \times  10^{42} ~\rm s^{-1} ~\MSUN^{-1}$, where we consider the X-ray band to span from 100 eV to 10 keV. Just as the ionizing ultraviolet radiation, the X-ray emission scales linearly with halo mass. We adopt a power-law spectral energy distribution (SED), i.e. $I_X(E) \propto E^{-\alpha}$, where $\alpha=1.2$ across the X-ray band. The $\dot N_X$ value for $f_X=1$ is in agreement with the SED of high-mass X-ray binaries in local star-forming galaxies for the 0.5--8 keV band \citep{2012MNRAS.419.2095M, 2019MNRAS.487.2785I}. We vary ${\log}_{10}$($f_X$) in the range [-3, 3].
    
    \item Minimum mass of UV emitting halos ($\MMIN$): This parameter determines the number density of UV emitting sources, as we assume that only dark matter halos with a mass larger than $\MMIN$ contribute to the ionizing photon budget. We vary  ${\log}_{10}$($\MMIN/\MSUN$) in the range [9, 12]. The lower value is determined by the lowest mass halos resolved in our N-body simulations. While the number density of dark matter halos in a $\Lambda$CDM cosmology is larger at the low mass end, star formation in halos with mass $\lesssim 10^9 ~\MSUN$ is expected to be significantly suppressed due to thermal feedback, SNe feedback, or inefficient gas accretion \citep[see e.g.,][]{1994ApJ...427...25S, Barkana2001, 2003MNRAS.339..312S, 2013MNRAS.432L..51S}. These lower mass halos are therefore expected to have a less significant impact on the 21-cm signal power spectrum.
    
    \item Minimum mass of X-ray emitting halos ($\MMINX$): This determines the number density of X-ray emitting halos, as we assume that only halos with a mass larger than $\MMINX$ host X-ray sources. Also for ${\log}_{10}$($\MMINX/\MSUN$) we explore the range [9, 12].
    
    \item Radio background efficiency ($A_r$): We assume the existence of a radio background\footnote{The existence of a radio background of astrophysical or cosmological origin in addition to the CMB is motivated by the results of ARCADE2 \citep{2011ApJ...734....5F} and LWA1 \citep{2018ApJ...858L...9D} observations, which show evidence of excess towards the Rayleigh-Jeans part of the spectrum.  In particular, the LWA1 measurement at frequency 40--80 MHz suggests a power-law dependence with spectral index -2.58$\pm$0.05 and a temperature of $603^{+102}_{-92}$ mK at 1.42 GHz. We note that the ARCADE2 and LWA1 measurements are debated, as the measurements might suffer from erroneous Galactic modelling \citep{2013ApJ...776...42S}. } in addition to the CMB. This parameter quantifies the effective brightness temperature of the total radio background as \citep{2019MNRAS.486.1763F,2020MNRAS.498.4178M}: 
    \begin{equation}
    \TEFF = T_{\gamma} \left[1+A_r \left(\frac{\nu_{\rm obs}}{78 ~\rm MHz} \right)^{-2.6} \right]. 
    \label{eq:trad}
    \end{equation}
     Here, $\TCMB(z)$ = $2.725~(1+z)$ K is the usual CMB temperature and $\nu_{\rm obs}$ is the observation frequency. For simplicity, we adopt a uniform radio background, but the presence of inhomogeneities could enhance the amplitude of the 21-cm signal power spectrum above our estimates \citep[][]{Reis2020Highredshift}. We vary $A_r$ in the range [0, 416]. The lower bounds correspond to no excess radio background (i.e. $\TEFF = T_{\gamma}$), while the upper bounds represent the upper limits of 0.603~K as measured by LWA1 \citep{2018ApJ...858L...9D}.
\end{itemize}

\begin{table}
\centering
\caption[]{Overview of the source parameters used in {\sc grizzly}  with their explored ranges. Note that the parameter space exploration is done on the log-spaced grids for each parameter.}
\small
\tabcolsep 1.3pt
\renewcommand\arraystretch{1.8}
   \begin{tabular}{c c c c c}
\hline
\makecell{Source \\ Parameters} & Description & Explored range  	 \\
\hline
$\zeta$ & Ionization efficiency   & [$10^{-3}, 10^{3}$]   &   	\\
$\MMIN$ & \makecell{Minimum mass of UV emitting halos}   & [$10^{9}  ~\MSUN, 10^{12} ~\MSUN$]        & 	\\
$\MMINX$ & \makecell{Minimum mass of X-ray emitting halos}   & [$10^{9} ~\MSUN, 10^{12} ~\MSUN$]   &  	\\
$\FX$ & X-ray heating efficiency   & [$10^{-3}$, $10^3$]   & 	\\
$A_r$ & \makecell{Excess radio background efficiency} & [0, 416]  & \\
\hline
\end{tabular}
\label{tab_source_param}
\end{table}


\begin{table}
\centering
\caption[]{List of the IGM parameters considered in this paper. We note that a heated region in the IGM has gas temperatures larger than $\TEFF$, i.e. $\TB \ge 0$.}
\small
\tabcolsep 2pt
\renewcommand\arraystretch{1.8}
   \begin{tabular}{c c c}
\hline
IGM Parameter & Description  &	 \\
\hline
$\AVXHII$ & Volume averaged ionized fraction       &	\\
$\AVTK$ (K) & \makecell{Average temperature of gas with $\XHII<0.5$}         &	\\
$\AVTB$ (mK) & \makecell{Mass averaged  $\TB$}         &	\\
$\FHEAT$ & \makecell{Volume fraction of regions hotter than $\TEFF$}   &\\
$\RPEAK$ ($\mpcbyh$) & \makecell{Size at which the PDF of the \\ heated regions size distribution peaks}         & 	\\
$\RFWHM$ ($\mpcbyh$) & \makecell{FWHM of the PDF of the \\ heated regions size distribution}       &	\\
\hline
\end{tabular}
\label{tab_igm_param}
\end{table}

\subsubsection{Simulating the 21-cm signal power spectrum}
\label{21cmsignal}
The 21-cm signal brightness temperature ($\TB$) can be written as \citep[see e.g,][]{madau1997, Furlanetto2006}, 
\begin{eqnarray}
 \TB (\mathbf{x}, z) \!  & = & \! 27 ~ x_{\rm HI} (\mathbf{x}, z) [1+\delta_{\rm B}(\mathbf{x}, z)]  \left(1-\frac{\TEFF(z)}{\TS(\mathbf{x}, z)} \right) \nonumber\\
&\times& \! \left(\frac{\OmegaB h^2}{0.023}\right) \left(\frac{0.15}{\Omegam h^2}\frac{1+z}{10}\right)^{1/2}  \left(\frac{H}{\mathrm{d}v_\mathrm{r}/\mathrm{d}r+H}\right)\,\rm{mK}.
\nonumber \\
\label{eq:brightnessT}
\end{eqnarray}
Here, $[1+\delta_{\rm B}(\mathbf{x}, z)]$ is the overdensity at position $\mathbf{x}$ and redshift $z$, extracted from the gridded density fields obtained from the N-body simulation described in Sect. \ref{sec:simul}. The quantities $\XHI$, $\TS$, $H$ and $\mathrm{d}v_\mathrm{r}/\mathrm{d}r$ are the neutral hydrogen fraction, the spin temperature of hydrogen in the IGM,  the Hubble parameter and the velocity gradient along the line of sight, respectively. Given a set of astrophysical source parameters, {\sc grizzly} generates $\XHI$ and gas temperature ($\TK$) maps. Here, we assume $\TS=\TK$ which is expected at the redshifts of interest due to the presence of a strong $\lya$ background \citep[e.g.][]{barkana05b, Pritchard12, 2021PhRvD.103h3025S}. We include the line-of-sight velocity field effects (so-called redshift space distortions, RSD) to the $\TB$ cubes using the cell movement method (or
Mesh-to-Mesh Real-to-Redshift-Space-Mapping scheme; \citealt{mao12, ghara15b, 2021MNRAS.506.3717R}). Finally, we estimate the dimensionless power spectrum of $\TB$, denoted as $\DTB(k)$\footnote{We note that in our previous studies  \citep{2020MNRAS.493.4728G, 2021MNRAS.503.4551G} we ignored the redshift space distortions, which results in underestimating the 21-cm signal power spectrum, especially when it is dominated by density fluctuations.}.

\subsubsection{Derived IGM parameters}
\label{sec_igm_param}
Similar to \citet{2020MNRAS.493.4728G, 2021MNRAS.503.4551G}, our main focus here is on the characterization of the IGM, as its properties are directly linked to the observed 21-cm signal. However, our framework is not completely independent of source models, as {\sc grizzly} uses a set of source parameters as input to simulate ionization and temperature fields, which are then used to determine a set of quantities related to the IGM. 

When considering the IGM parameters, we define `heated regions' as those with $\TK>\TEFF$. To derive the size distribution of these regions we use the mean free path (MFP) method from \citet{2007ApJ...669..663M}, see \cite{2020MNRAS.493.4728G} for more details. We then consider the following derived IGM parameters/quantities\footnote{Unlike in \citet{2020MNRAS.493.4728G}, here we do not consider separate probability distribution functions (PDFs) for the \HII regions. In the absence of X-ray heating, the \HII regions are equivalent to the heated regions, as the gas temperature within them is $\sim 10^4$K, i.e. much larger than $\TEFF$.} for each redshift (see also Table \ref{tab_igm_param}):

\begin{itemize}
    \item $\AVXHII$: Volume averaged ionized fraction.
    \item $\AVTK$ (K): Volume averaged gas temperature of neutral regions, where we consider the gas to be neutral if $\XHII< 0.5$. We note that choosing a different value would change the estimate of $\AVTK$. In general, a higher $\XHII$ threshold results in a larger value of $\AVTK$. 
    \item $\AVTB$ (mK):  Mass averaged differential brightness temperature.
    \item $\FHEAT$: Volume fraction of heated regions ($\TK>\TEFF$). 
    \item $\RPEAK$ ($\mpcbyh$): Characteristic size of the heated regions, corresponding to the size at which the distribution of heated regions size peaks.
    \item $\RFWHM$ ($\mpcbyh$): Full width at half maximum (FWHM) of the PDF of the size distribution of heated regions.
\end{itemize}

%

\subsubsection{Bayesian inference}
\label{sec:like}
To explore the source parameter space as given in Table \ref{tab_source_param}, we use the publicly available Monte Carlo Markov Chain (MCMC) sampler code {\sc cobaya\footnote{\url{https://cobaya.readthedocs.io/en/latest/index.html}}}\citep{2002PhRvD..66j3511L}.  The likelihood of a set of parameters $\bm{\theta}$ to be excluded at $z$ for an upper limit of the power spectrum $\DTBLOFARMEAN (k_i, z) \pm \DTBLOFARSIGMA (k_i, z)$,  can be defined as \citep[see][ for details]{2020MNRAS.493.4728G}
\begin{equation}
    \mathcal{L}_{\mathrm{ex, single}-z}(\bm{\theta}, z) = 1- \prod_{i} \frac{1}{2}\left[ 1 + \mathrm{erf}\left(\frac{\DTBLOFARMEAN (k_i, z)- \DTBMODELMEAN(k_i, \bm{\theta}, z)}{\sqrt{2} \sigma (k_i, z)}\right) \right].
    \label{equ:like}
\end{equation}
Here, $\DTBMODELMEAN(k, \bm{\theta}, z)$ is the model power spectrum for a set of parameters $\bm{\theta}$, while $\sigma$ is the error which includes contributions from the observational error ($\DTBLOFARSIGMA$), as well as from the power spectrum modelling error ($\DTBMODELSIGMA$), i.e. $\sigma = \sqrt{ \DTBLOFARSIGMASQ + \DTBMODELSIGMASQ}$. 

Unlike our previous inference studies, here we also consider the combination of the three upper limits into a joint likelihood
\begin{equation}
    \mathcal{L}_{\mathrm{ex, joint}-z}(\bm{\theta}) = 1- \prod_{i,j} \frac{1}{2}\left[ 1 + \mathrm{erf}\left(\frac{\DTBLOFARMEAN (k_i, z_j)- \DTBMODELMEAN(k_i, \bm{\theta}, z_j)}{\sqrt{2} \sigma (k_i, z_j)}\right) \right].
    \label{equ:likejoint}
\end{equation}
We note that multiple upper limits can be combined into a single likelihood only when the redshift evolution of the source parameters is well understood. 
As here we assume the source parameters to be redshift-independent, some caution should be applied in the interpretation of the joint analysis. In the future, we plan to extend it by employing the code {\sc polar} \citep{2023MNRAS.522.3284M, 2024arXiv241011620A}, which includes a redshift evolution in the modelling of the source parameters.

Similar to \citet{2021MNRAS.503.4551G}, our framework does not directly use {\sc grizzly} in the MCMC analysis. 
Instead, we first employ {\sc grizzly} to generate $10^5$ sets of power spectra and IGM parameters at each of the three redshifts by sampling the 5D source parameters space\footnote{Each source parameter, $s_{\rm i}$, is sampled by linearly binning the range [${\log}_{\rm 10}(s_{\rm i,min})$, ${\log}_{\rm 10}(s_{\rm i,max})$].}. Then, in the MCMC analysis we employ a linear interpolation scheme \citep{weiser1988note, Virtanen2020scipy} over the 5D source parameter space, which uses the previously generated power spectra and IGM parameters. We also consider a conservative 30\% modelling error, $\DTBMODELSIGMA (k_i, z) = 0.3\times \DTBMODELMEAN(k_i, \bm{\theta}, z)$. This includes the inherent modelling errors due to the approximate nature of the {\sc grizzly} algorithm\footnote{A comparison study of {\sc grizzly} with the 3D radiative transfer scheme C$^2${\sc ray} for different reionization scenarios assuming $\TS\gg\TCMB$ showed an excellent match of the power spectra, with less than 10 per cent difference \citep{ghara18}. However, the level of agreement across the entire parameter space is not known. Also, we do not have robust accuracy estimates of {\sc grizzly} for scenarios with $\TS$ fluctuations. Thus we use a conservative modelling error which also covers other uncertainties such as an interpolation error in the MCMC analysis. We note that the chosen modelling error is still significantly smaller than the $1\sigma$ error on the LOFAR upper limits at the large scales.}, as well as the errors introduced by the interpolation scheme. To run the MCMC on the source parameters we use 10 walkers and $10^6$ steps, and ensure that all the MCMC chains converge well before the final step. We adopt a uniform prior in $\log$-scale on the explored parameter ranges (see Table \ref{tab_source_param}). {\sc cobaya} produces the list of corresponding derived IGM parameters using the same linear interpolation scheme. 

As the upper limits towards the largest $k$ values exceed the values for all of the power spectra in our simulation database (see next section), they do not yield any constraints. Therefore, only the three smallest $k$-bins (see Table \ref{tab_obs}) are used in the MCMC analysis to estimate $\mathcal{L}_{\mathrm{ex, single}-z}(\bm{\theta}, z)$ and $\mathcal{L}_{\mathrm{ex, joint}-z}(\bm{\theta})$.

In this analysis we also use a prior on $\AVXHII$, derived from the measured Thomson scattering optical depth towards the CMB. For each of the three redshifts $z_i$ we assume a stepwise reionization history, in which the Universe is neutral until $z_i$, partially ionized at a level ($\AVXHIIMAX$) until redshift 6, and fully ionized after that\footnote{The exact redshift of the end of reionization ($z_{\rm end}$) is debated. While the Gunn-Peterson trough in high$-z$ quasar spectra suggests $z_{\rm end}\approx 6$ \citep[e.g.][]{Fan06b, Mortlock11, Venemans15,  2018Natur.553..473B},  $\lya$ forest observations at lower redshifts suggest a late reionization \citep[see, e.g.,][]{2018ApJ...863...92B, 2018ApJ...864...53E}}. The value $\AVXHIIMAX$ follows from assuming the $1\sigma ~\tau$ measurement, i.e., $\tau = 0.061$ for $\tau = 0.054\pm 0.007$  \citep{2020A&A...641A...6P}. It represents the largest possible of value of $\AVXHII$ at redshift $z_i$ consistent with the $\tau$ measurements. We obtain values of $\AVXHIIMAX=1$,  0.79 and 0.57 at $z=8.3$, 9.1 and 10.1, respectively. The same approach was used in \citet{2020MNRAS.493.4728G, 2021MNRAS.503.4551G}. We note that in a realistic reionization scenario, the evolution of $\AVXHII$ is expected to be gradual.

It should be noted that the Bayesian analysis explores the source parameters rather than the IGM parameters, as our framework uses the former as input. The source parameter values obtained from the MCMC chains are then used to generate lists of corresponding IGM parameter values. Finally, these lists are employed to obtain the posterior distribution of the IGM parameters. Because of this procedure, the constraint on the IGM parameters might be significantly affected by the priors on the source parameters. We discuss this in Appendix \ref{res:appe2}.

\begin{figure}
\begin{center}
\includegraphics[scale=0.5]{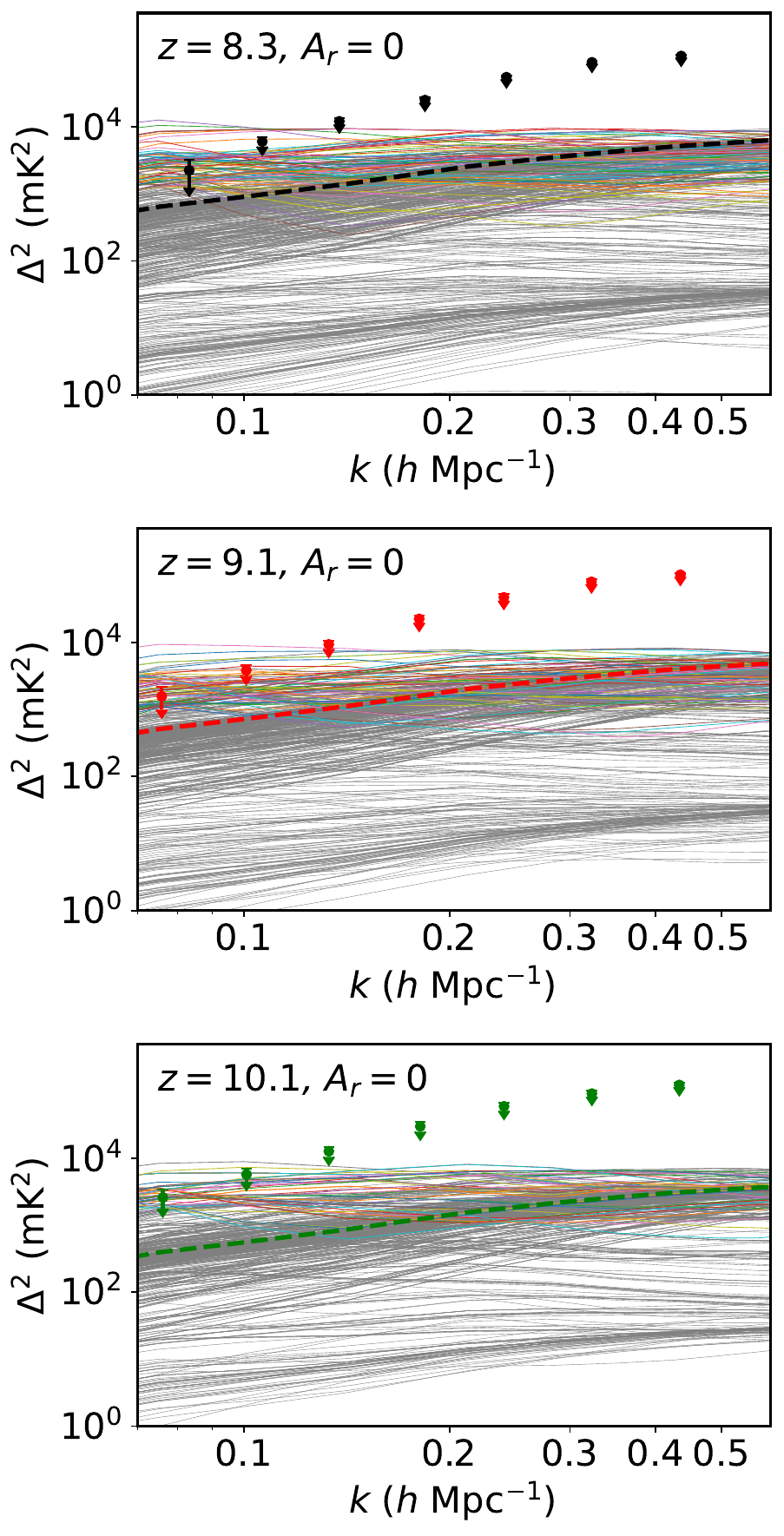}
    \caption{A set of 1000  power spectra randomly chosen out of the $ 10^5$ simulated ones. Panels from top to bottom refer to $z=8.3$, 9.1 and 10.1, respectively.
    These power spectra correspond to the scenario with no additional radio background other than the CMB ($A_r = 0$). The down arrow points refer to the recent LOFAR 1$\sigma$ upper limit ($\DTBLOFARMEAN (k, z) \pm \DTBLOFARSIGMA (k, z)$) from \citet{lofar2024}. As a reference, the dashed line corresponds to the power spectrum for a completely neutral IGM with no X-ray heating, and with $\TS=\TK$. The coloured power spectra are those with a value larger than  $\DTBLOFARMEAN (k, z) - \DTBLOFARSIGMA (k, z)$ in at least one $k-$bin, i.e. they have a high probability of being ruled out, while those in grey have a low exclusion probability. }
   \label{image_psts_z}
\end{center}
\end{figure}

\begin{figure}
\begin{center}
\includegraphics[scale=0.45]{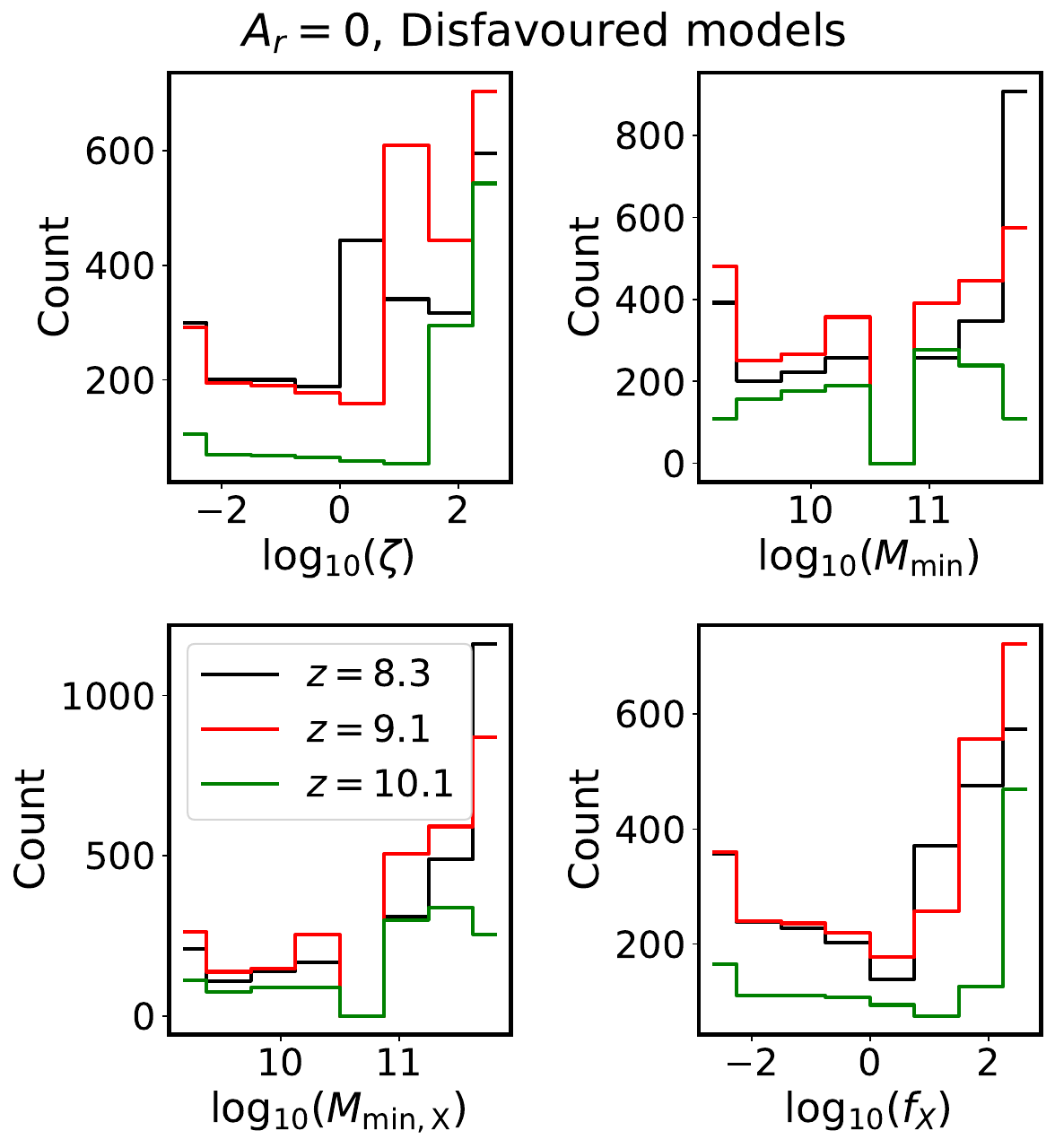}
    \caption{Histogram of the disfavoured values of the source parameters for the scenario with $A_r=0$. These correspond to the models with power spectra values larger than  $\DTBLOFARMEAN (k, z) - \DTBLOFARSIGMA (k, z)$ in at least one $k-$bin, i.e. which have a high probability of being ruled out by the recent LOFAR 1$\sigma$ upper limits ($\DTBLOFARMEAN (k, z) \pm \DTBLOFARSIGMA (k, z)$) from \citet{lofar2024}.}
   \label{image_hist_z}
\end{center}
\end{figure}

\begin{figure}
\begin{center}
\includegraphics[scale=0.45]{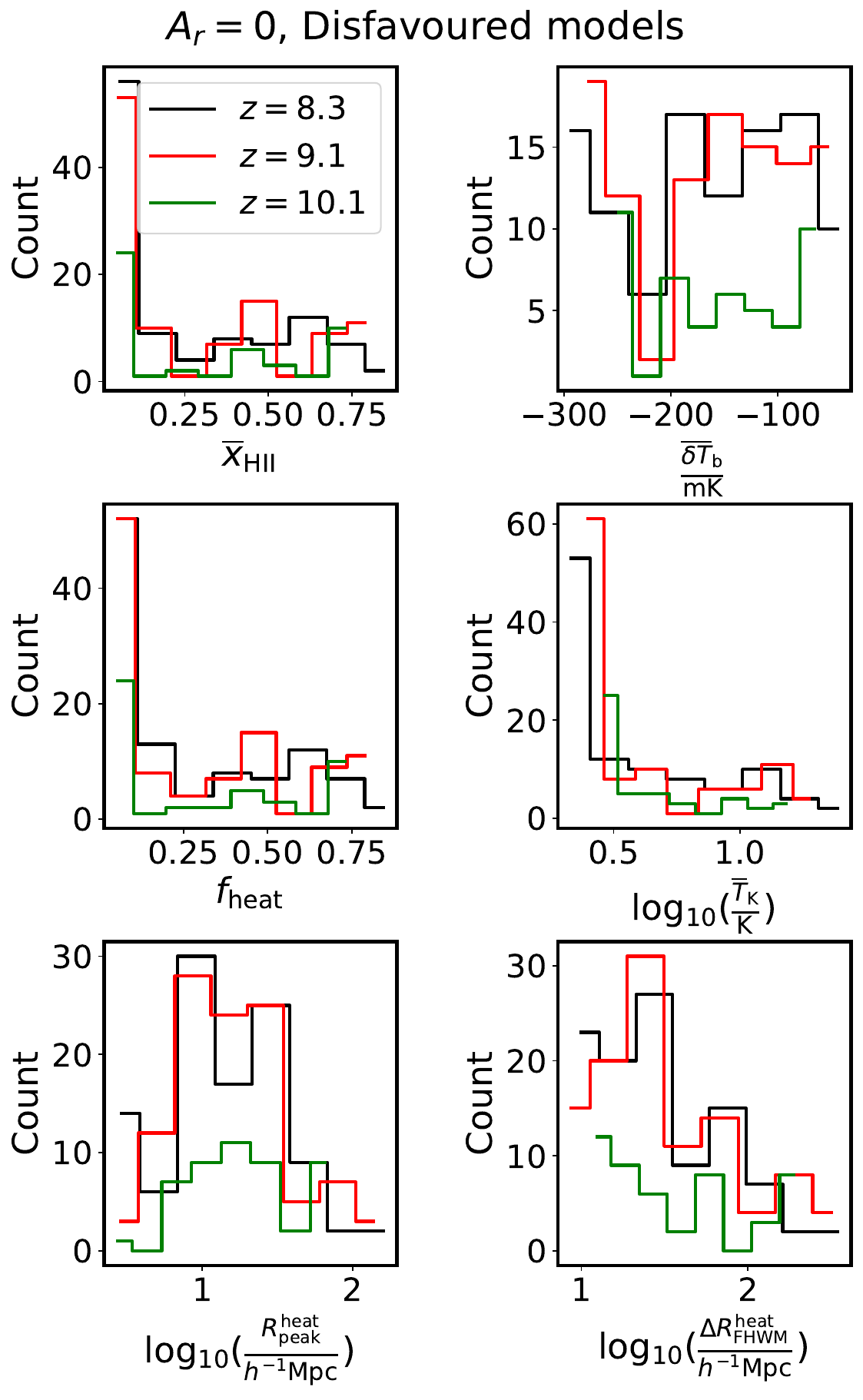}
    \caption{Histogram of the disfavoured values of the IGM parameters for the scenario with $A_r=0$. These correspond to the same set of models considered in Fig. \ref{image_hist_z}.}
   \label{image_histIGM_z}
\end{center}
\end{figure}

\section{Results}
\label{sec:results}
Before describing the quantitative analysis of our results, we present in Fig. \ref{image_psts_z}, as an illustration, a subset of the 21-cm signal power spectra with $A_r = 0$ at $z=8.3$, 9.1 and 10.1. We find that at each redshift some of the power spectra (indicated with coloured lines) have values larger than $\DTBLOFARMEAN (k, z) - \DTBLOFARSIGMA (k, z)$ from the recent LOFAR 1$\sigma$ upper limits from \citet[][indicated by the down arrow points]{lofar2024} in at least one $k-$bin. For this reason, they have a high probability of being ruled out. Overall, we find that 10, 13 and 5 per cent of our simulated power spectra fall into this category at $z=8.3$, 9.1 and 10.1, respectively. Here it should be kept in mind that these fractions depend on the priors chosen for the source parameters. On the other hand, the power spectra in grey have a low exclusion probability as they remain below that limit. The thick dashed lines correspond to the power spectra of a completely neutral and unheated IGM, obtained assuming constant values of $\XHI=1$ and $\TS=\TK=0.021\times (1+z)^2 ~\mathrm K$.  We note that the observed upper limits are still unable to rule out this scenario. We estimate that for $z=9.1$ at least 2-3 times more integration time would be required to achieve that.

As a further illustration, Fig. \ref{image_hist_z} shows the distribution of source parameters for models with a high probability of being ruled out, i.e. which have a value larger than  $\DTBLOFARMEAN (k, z) - \DTBLOFARSIGMA (k, z)$ in at least one $k-$bin. The vertical axes represent the number of disfavoured models out of the total of $10^5$. We only consider the $A_r=0$ scenario and each source parameter range has been logarithmically binned into eight bins. We note that a significant fraction of these disfavoured models has high values of $\MMIN$ and $\MMINX$, indicating that they represent scenarios in which the ionized or heated regions are rare. A large fraction of models also has high values of $\zeta$ and $\FX$, suggesting that the rare ionized and heated regions are large in size.

\begin{figure}
\begin{center}
\includegraphics[scale=0.9]{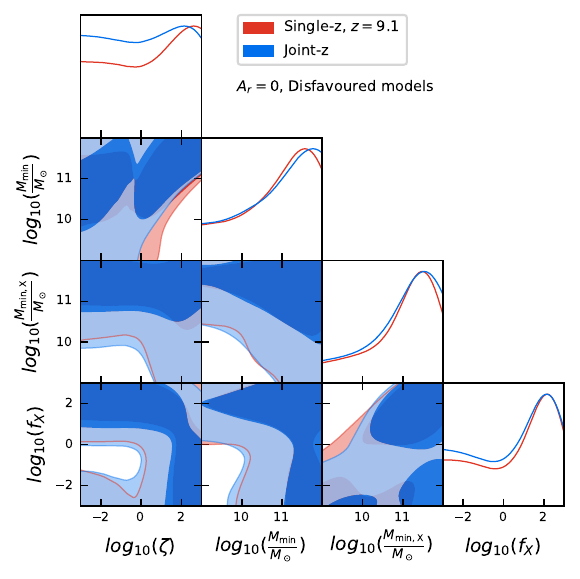}
    \caption{Posterior distribution of the source parameters of the models which are disfavoured by the LOFAR upper limits from \citet{lofar2024} at $z= 9.1$ for the $A_r=0$ scenario. Red refers to the case when each redshift is considered separately ({\it Single-z}) at $z=$9.1, while blue represents the joint MCMC analysis results ({\it Joint-z}). The contour levels in the two-dimensional contour plots refer to the $1\sigma$ and $2\sigma$ credible intervals of the models disfavoured by the LOFAR upper limit at this redshift. The diagonal panels represent the corresponding marginalized probability distributions of each parameter. }
   \label{image_sourceparam_Ar0}
\end{center}
\end{figure}

\begin{figure}
\begin{center}
\includegraphics[scale=1.]{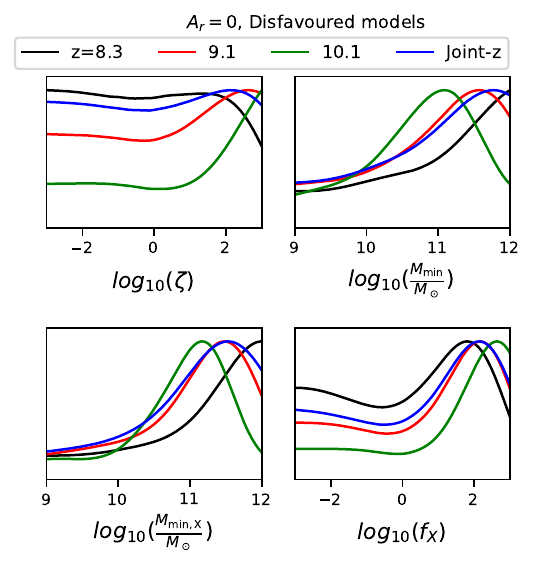}
    \caption{Marginalized probability distribution of the source parameters of the models which are disfavoured by the LOFAR upper limits from \citet{lofar2024} at $z= 9.1$ for the $A_r=0$ scenario. The black, red and green curves refer to the {\it Single-z} case at $z=8.3$, 9.1 and 10.1 respectively, while the blue curves are for the {\it Joint-z} analysis case. }
   \label{image_source1dall_Ar0}
\end{center}
\end{figure} 

In Fig. \ref{image_histIGM_z}, we present the distribution of IGM parameters for the same set of models considered in Fig. \ref{image_hist_z}. A significant number of disfavoured models at all redshifts have $\AVXHII$ and $\FHEAT$ smaller than $\sim 0.6$. The characteristic sizes in these models are several tens of $\mpcbyh$, indicating that the ionized or heated regions are both rare and large. At the same time, the volume occupied by them remains less than 50 per cent of the total volume. We note that a large number of disfavoured models with $\AVXHII\lesssim0.1$, $\FHEAT\lesssim0.1$ and $\AVTK\lesssim$4 K are strongly influenced by the priors chosen on the source parameters (see Appendix \ref{res:appe2} for details).

In the following sections, we present the results of our MCMC analysis. Following the same approach adopted in \citet{2021MNRAS.503.4551G}, we first perform the MCMC using the likelihood $\mathcal{L}_{\mathrm{ex, single}-z}(\bm{\theta}, z)$ as given in Eq. (\ref{equ:like}), which accounts for upper limits at a single redshift only. We denote this as `{\it Single-z}' analysis. We also run an MCMC analysis where we use the joint likelihood $\mathcal{L}_{\mathrm{ex, joint}-z}(\bm{\theta})$ as shown in Eq. (\ref{equ:likejoint}), which combines all the upper-limit results. We denote this as `{\it Joint-z}' analysis. We note again that as our source parameters are assumed to be redshift independent,  the results of the {\it Joint-z} analysis should be interpreted with some caution. We will first consider the case of no additional radio background, followed by the case in which we allow for an additional radio background. Finally we describe how the inclusion of upper limits on the power spectrum from other 21-cm experiments impacts the results.

\begin{figure}
\begin{center}
\includegraphics[scale=1.2]{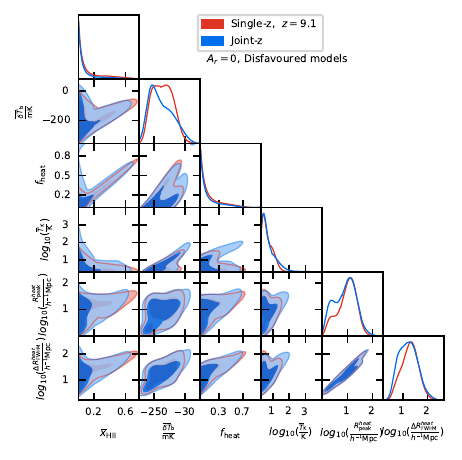}
    \caption{Posterior distribution of the IGM parameters of the models which are disfavoured by the LOFAR upper limits from \citet{lofar2024} at $z= 9.1$ for the $A_r=0$ scenario. Red refers to the case when $z= 9.1$ is considered separately ({\it Single-z}), while blue represents the joint redshift analysis obtained at $z=$9.1 ({\it Joint-z}). The contour levels in the two-dimensional contour plots refer to the $1\sigma$ and $2\sigma$ credible intervals of the models disfavoured by the LOFAR upper limit at this redshift. The diagonal panels represent the corresponding marginalized probability distributions of each parameter. }
   \label{image_igmparam_Ar0}
\end{center}
\end{figure}

\begin{table}
\centering
\caption[]{Constraints on the source parameters for the $A_r=0$ scenario, derived using the LOFAR upper limits from \citet{lofar2024}. The first column shows the list of source parameters, the following three constraints obtained from the {\it Single-z} analysis at three redshifts, and the last one those obtained with the {\it Joint-z} analysis. All these are disfavoured limits at 68$\%$ credible intervals level. The `X' sign means that the credible intervals span the entire prior range of a parameter. The 95 per cent disfavoured credible intervals span the entire prior ranges for all cases and thus are not shown here. }  
\small
\tabcolsep 2.6pt
\renewcommand\arraystretch{2.4}
   \begin{tabular}{| c | c c c c c|}
\hline

\makecell{Source  \\ Parameters }  & \makecell{$z=8.3$\\({\it Single-z})} & \makecell{$z=9.1$\\({\it Single-z})} & \makecell{$z=10.1$\\({\it Single-z})} & {\it Joint-z} & \\

\hline
\multirow{1}{4em}{ ${\log}_{\rm 10}(\zeta)$ }
& [-3, 0.94] & X & [-0.53, 3]  & X & \\
\cline{2-6}

\hline

\multirow{1}{4em}{${\log}_{\rm 10}(\frac{\MMIN}{\MSUN})$ }

&  [10.45, 12] & [10.49, 12] & [10.23, 11.7] &  [10.45, 12] &\\ 
\cline{2-6}

\hline
\multirow{1}{4em}{${\log}_{\rm 10}(\frac{\MMINX}{\MSUN})$ }

&  [10.78, 12.] & [10.76, 12] & [10.5, 11.7]  & [10.7, 12] & \\ 
\cline{2-6}

\hline

\multirow{1}{4em}{${\log}_{\rm 10}(f_X)$ }

&  [-3, 1.54] & X & [0.6, 3]  & X & \\ 
\cline{2-6}

\hline

\end{tabular}
\label{tab_mcmc_sourceparamAr0}
\end{table}

\subsection{Analysis without excess radio background ($A_r=0$)}
\label{sec:noRBG}
For $A_r=0$ our parameter space is 4D, defined by $\zeta$, $\MMIN$, $\MMINX$ and $f_X$. This scenario corresponds to one in which the only radio background at 1420 MHz is the CMB. 

Figure \ref{image_sourceparam_Ar0} shows the posterior distribution of the {\sc grizzly} source parameter space at $z=9.1$. Here we highlight that the likelihood used in the MCMC represents the exclusion probability of a model. We also note that the 68 and 95 per cent credible interval contours on the two-dimensional contour panels have remained open, indicating that the constraints on the source parameters depend on the chosen prior ranges. We find that the 68 per cent disfavoured credible intervals\footnote{A credible interval is a parameter range with a certain probability (e.g., 68 per cent) of containing the true value of the parameter. Our constraints on the source and IGM  parameters do not represent models ruled out with a high significance, but rather models with a high probability to be excluded. These interval limits are also highly dependent on the prior chosen on the source parameter ranges.} from the {\it Single-z} analysis at $z=9.1$ are $\MMIN \gtrsim 3\times 10^{10} ~\MSUN$ and $\MMINX \gtrsim 6\times 10^{10} ~\MSUN$, while the limits for $\zeta$ and  $f_X$ span the entire parameter ranges. The 95 per cent disfavoured credible intervals of the source parameters span the entire parameter ranges for both the {\it Single-z} and {\it Joint-z} analysis. The 68 per cent credible intervals for the {\it Joint-z} analysis case are similar to those of the {\it Single-z} case. One reason behind the similarity could be that the  {\it Joint-z} analysis results are likely to be biased by the strongest upper limit results, which are at $z=9.1$. 

The marginalized probability distributions of the disfavoured source parameter values at all redshifts are shown in Fig. \ref{image_source1dall_Ar0}. The related credible intervals are listed in Table \ref{tab_mcmc_sourceparamAr0} for both the ${\it Single-z}$ and ${\it Joint-z}$ cases. The credible intervals for $\MMIN$ and $\MMINX$ for the three {\it Single-z} cases are qualitatively consistent with each other, as they disfavour models at the highest end of the prior range, i.e. ${\log}_{\rm 10}(\MMIN)$ or ${\log}_{\rm 10}(\MMINX)$ $\gtrsim 10.5~\MSUN$. This suggests that models associated with a low number density of massive UV and X-ray emitting sources are disfavoured. The limits for $\zeta$, on the other hand, differ among the various redshifts. For $z=10.1$ models towards the higher end of our prior interval are disfavoured, but for $z=8.3$ it is rather models at the lower end of our prior interval which are found to be less likely. The same trend is seen for the $f_X$ parameter. This can be understood from the difference in the number density of halos between these two redshifts. Scenarios more likely to be disfavoured require a high contrast, so the ionized/heated regions should not be too small/few or too large/many. With a smaller halo density at $z=10.1$, relatively large efficiencies are needed to produce a large contrast population of ionized/heated regions, but at the larger halo density for $z=8.3$ such high efficiencies would ionize/heat too much of the IGM, yielding a low contrast, and instead lower efficiencies are needed to create a large contrast distribution of ionized/heated regions.

\begin{table}
\centering
\caption[]{Constraints on the IGM parameters for the $A_r=0$ scenario, derived using the LOFAR upper limits from \citet{lofar2024}.
The first column shows the list of IGM parameters derived from both the {\it Single-z} and {\it Joint-z} analysis, the second refers to the credible intervals of the disfavoured models, while the last three to the constraints obtained from the analysis at three redshifts.} 
\small
\tabcolsep 3pt
\renewcommand\arraystretch{2.}
   \begin{tabular}{| c | c | c c c c |}
\hline

\makecell{IGM  \\ Parameters \\(MCMC  \\ Model)}  & \makecell{Credible  \\ intervals}  & $z=8.3$ & $z=9.1$ & $z=10.1$ &   \\

\hline
\multirow{2}{4em}{ $\AVXHII$ ({\it Single-z})}
& 68$\%$ & [0, 0.15] & [0, 0.13] & [0, 0.2]  & \\
\cline{2-6}

& 95$\%$ & [0, 0.57] & [0, 0.55] & [0, 0.47] & \\
\cline{1-6}
\multirow{2}{4em}{$\AVXHII$ ({\it Joint-z})}
&  68$\%$ & [0, 0.24] & [0, 0.06] & [0, 0.01]  & \\
\cline{2-6}

& 95$\%$ & [0, 0.85] & [0, 0.46] & [0, 0.17] & \\

\hline

\multirow{2}{4em}{$\FHEAT$ ({\it Single-z})}

& 68$\%$ & [0, 0.21] & [0, 0.18] & [0, 0.22] & \\ 
\cline{2-6}

& 95$\%$ & [0, 1] & [0, 0.57] & [0, 0.47] & \\
\cline{1-6}
\multirow{2}{4em}{$\FHEAT$ ({\it Joint-z})}
&  68$\%$ & [0, 0.49] & [0, 0.15] & [0, 0.05] & \\ 
\cline{2-6}

& 95$\%$ & [0, 0.9] & [0, 0.54] & [0, 0.21] & \\
\hline
\multirow{2}{4em}{ $\AVTK$ (K) ({\it Single-z})}

& 68$\%$ & [1.8, 7.7] & [2.2, 7.3] & [2.6, 12.8]  & \\ 
\cline{2-6}

& 95$\%$ & [1.8, 50] & [2.2, 21] & [2.6, 22]  & \\
\cline{1-6}
\multirow{2}{4em}{ $\AVTK$ (K) ({\it Joint-z})}
& 68$\%$ & [1.8, 28.5] & [2.2, 7.2] & [2.6, 5]  & \\ 
\cline{2-6}

& 95$\%$ & [1.8, 977] & [2.2, 35.6] & [2.6, 15]  & \\

\hline

\multirow{2}{4em}{$\AVTB$ (mK) ({\it Single-z})}

& 68$\%$ & [-300, -124] & [-276,-123] & [-222, -113]  & \\ 
\cline{2-6}

& 95$\%$ & [-345, -23] & [-317, -61] & [-283, -80] & \\
\cline{1-6}
\multirow{2}{4em}{$\AVTB$ (mK) ({\it Joint-z})}
& 68$\%$ & [-264, -4] & [-308,-141] & [-308, -220]   & \\ 
\cline{2-6}

& 95$\%$ &  [-313, 10] & [-337, -36] & [-289., -116] & \\

\hline
\multirow{2}{4em}{$\RPEAK$ ($\mpcbyh$) ({\it Single-z})}

& 68$\%$ & [0, 26.3] & [0, 28.1] & [0, 20.6]  & \\ 
\cline{2-6}

& 95$\%$ & [0, 56.4.] & [0, 39.5] & [0, 35.5] & \\
\cline{1-6}
\multirow{2}{4em}{$\RPEAK$ ($\mpcbyh$) ({\it Joint-z})}
& 68$\%$ & [0, 68] & [0, 25] & [0, 14]  & \\ 
\cline{2-6}

& 95$\%$ & [0, 205] & [0, 41] & [0, 23] & \\

\hline
\multirow{2}{4em}{$\RFWHM$ ($\mpcbyh$) ({\it Single-z})}

& 68$\%$ & [6, 47] & [9, 46] & [11,38]  & \\ 
\cline{2-6}

& 95$\%$ & [3, 118] & [4, 81] & [3.5, 79] & \\
\cline{1-6}
\multirow{2}{4em}{$\RFWHM$ ($\mpcbyh$) ({\it Joint-z})}
& 68$\%$ & [0, 135] & [0, 40] & [0, 21]  & \\ 
\cline{2-6}

& 95$\%$ & [0, 331] & [0, 85] & [0, 33] & \\

\hline

\end{tabular}
\label{tab_mcmc_igmAr0}
\end{table}

Indeed, as also seen in our previous studies \citep{2020MNRAS.493.4728G, 2021MNRAS.503.4551G,2024A&A...687A.252G} only a few IGM states can produce very large amplitudes for the large-scale power spectrum. One of these is a patchy reionization model with cold neutral regions. Such a model requires reionization by highly luminous UV emitting sources (large value of $\zeta$) with a low number density  (large value of $\MMIN$), combined with a small value of $f_X$ to keep the neutral regions cool. In such a scenario, the large-scale $\DTB$ is dominated by $\XHII$ fluctuations and reaches the largest amplitude for $\AVXHI \sim 0.5$ \citep[see, e.g.,][]{2024A&A...687A.252G}. A second possibility is a patchy heating scenario, which requires rare and bright X-ray sources, implying large values for both $f_X$ and $\MMIN$.

Figure \ref{image_igmparam_Ar0} shows the posterior distributions of the disfavoured IGM parameter values at $z=9.1$. The corresponding credible intervals are listed in Table \ref{tab_mcmc_igmAr0}. The disfavoured models have $\AVXHII \lesssim 0.13 ~(0.55)$,  $\AVTK \lesssim 7.3 ~(21)$ K, $\FHEAT \lesssim 0.18 ~(0.57)$, $-276 ~(-317) \lesssim \AVTB \lesssim -123 ~(-61)$  mK, $ \RPEAK \lesssim 28 ~(40) ~\mpcbyh$ and $ \RFWHM\lesssim 46 ~(81)~\mpcbyh$ at 68 (95) per cent credible intervals for the {\it Single-z} analysis case.  The PDF of $\AVXHII$ indicates that a significant fraction of the disfavoured models are nearly fully neutral ones, where the large-scale power spectrum is mainly determined by fluctuations in $\TS$. As previously mentioned, the derived constraints on the IGM parameters might be significantly affected by the chosen priors on the source parameters (see Appendix \ref{res:appe2} for details).

Table \ref{tab_mcmc_igmAr0} also includes the results on the IGM parameters obtained from the MCMC analysis performed at the other two redshifts. The {\it Single-z} analysis gives fairly similar results at all redshifts, for both the 68 and 95 per cent credible intervals levels. Indeed, the disfavoured IGM states are those which cause a high contrast in the $\TB$ maps, producing large $\DTB$ values at scales $0.076 \hbympc \lesssim k\lesssim 0.133 \hbympc$.
Unlike the {\it Single-z} case, in the {\it Joint-z} analysis the IGM parameter's disfavoured credible intervals show a consistent evolution with redshift. For example, the 95 per cent credible intervals limits on $\AVXHII$ are $\lesssim 0.85$, $\lesssim 0.46$ and $\lesssim 0.17$ at $z=8.3$, 9.1 and 10.1, respectively. The main reason behind such evolution is that in this case the disfavoured limits of $\MMIN$ and $\MMINX$ are basically the same at all redshifts, while the number density of dark matter halos contributing to ionization/heating increases towards lower redshift for a fixed $\MMIN$ and $\MMINX$. This causes a more efficient ionization and heating process towards the lower redshift for disfavoured models from the {\it Joint-z} analysis.

\begin{table}
\caption[]{Constraints on the source parameters for the {\it Varying} $A_r$ scenario, derived using the LOFAR upper limits from \citet{lofar2024}. The first column shows the list of source parameters, the following three constraints obtained from the {\it Single-z} analysis at three redshifts, and the last one those obtained with the {\it Joint-z} analysis at $z=9.1$. All these are disfavoured limits at 68$\%$ credible intervals level. The `X' sign means that the credible intervals span the entire prior range of a parameter. The 95 per cent disfavoured credible intervals span the entire prior ranges for all four cases.} 
\small
\tabcolsep 2.6pt
\renewcommand\arraystretch{2.4}
   \begin{tabular}{| c | c c c c c|}
\hline

\makecell{Source  \\ Parameters }    & \makecell{$z=8.3$\\({\it Single-z})} & \makecell{$z=9.1$\\({\it Single-z})} & \makecell{$z=10.1$\\({\it Single-z})} & {\it Joint-z} & \\

\hline
\multirow{1}{4em}{ ${\log}_{\rm 10}(\zeta)$ }
 & [-3, 0.08] & [-3, 0.4] & [-3,0.61]  & [-3, 0.36] & \\
\cline{2-6}

\hline

\multirow{1}{4em}{${\log}_{\rm 10}(\frac{\MMIN}{\MSUN})$ }

 & [10.25, 12] & [10.23, 12] & [10.2, 11.7] &  [10.23, 12] &\\ 
\cline{2-6}

\hline
\multirow{1}{4em}{${\log}_{\rm 10}(\frac{\MMINX}{\MSUN})$}

 & [10.27, 12.] & [10.2, 12] & X  & [10.2, 12] & \\ 
\cline{2-6}

\hline

\multirow{1}{4em}{${\log}_{\rm 10}(f_X)$ }

 & [-3, 0.46] & [-3, 0.72] & [-3, 0.98]  & [-3, 0.85] & \\ 
\cline{2-6}

\hline
\multirow{1}{4em}{${\log}_{\rm 10}(A_r)$ }

 & [0.9, 2.62] & [0.66, 2.62] &[0.86, 2.62]  & [0.42, 2.62] & \\ 
\cline{2-6}

\hline

\end{tabular}
\label{tab_mcmc_sourceparamAr}
\end{table}


\subsection{Analysis with excess radio background ({\it Varying} $A_r$)}
\label{sec:RBG}
Next, we present the results of the MCMC analysis for the 5D source parameter space, i.e. $\zeta$, $M_{\rm min}$, $M_{\rm min, X}$, $f_X$ and $A_r$, the same as above with the addition of the excess radio background. Table \ref{tab_source_param} lists the priors on these parameters.

\begin{figure}
\begin{center}
\includegraphics[scale=0.9]{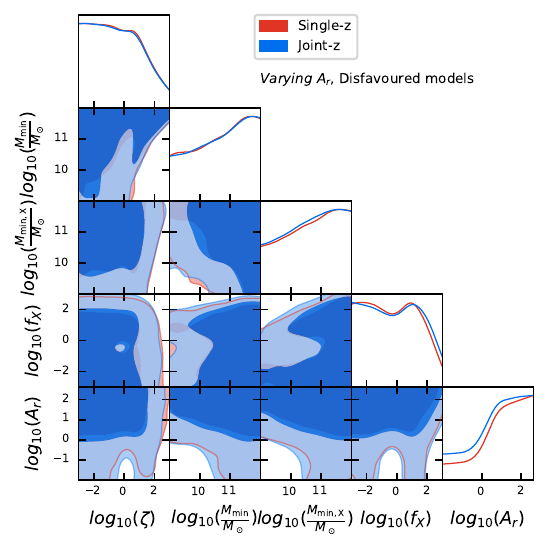}
    \caption{Posterior distribution of the source parameters of the models which are disfavoured by the LOFAR upper limits from \citet{lofar2024} at $z= 9.1$ for the {\it Varying} $A_r$ scenario. Red refers to the case when each redshift is considered separately ({\it Single-z}) at $z=$9.1, while blue represents the joint MCMC analysis results ({\it Joint-z}). The contour levels in the two-dimensional contour plots refer to the $1\sigma$ and $2\sigma$ credible intervals of the disfavoured models. The diagonal panels represent the corresponding marginalized probability distributions of each parameter. }
   \label{image_sourceparam_Ar}
\end{center}
\end{figure} 

Figure \ref{image_sourceparam_Ar} shows the posterior distribution of the five parameters for the disfavoured models at $z=9.1$. For the {\it Single-z} approach, the 68 per cent disfavoured credible intervals limits are  $\zeta \lesssim 2.4$, $\MMIN \gtrsim 1.7\times 10^{10} ~\MSUN$, $\MMINX \gtrsim 1.6\times 10^{10} ~\MSUN$, $f_X \lesssim 5.2$ and $A_r \gtrsim 4.6$ (see Table \ref{tab_mcmc_sourceparamAr}). The 95 per cent disfavoured credible intervals of the source parameters span the entire parameter ranges for all three {\it Single-z} and {\it Joint-z} analyses. The physical reasons behind the high amplitude of the large-scale power spectrum are the same as those discussed in the previous section. Additionally, a higher value of $A_r$ increases the signal strength, and thus the power spectrum amplitude. We note that the effect of $A_r$ is model-dependent. In fact, as for $A_r=0$, also in this scenario the signal from fully ionized gas is zero because $\TB=0$, while the strength of the signal (in absorption and/or emission) from neutral or partially ionized gas increases for larger values of $A_r$. This causes a different enhancement in the power spectrum amplitude between a scenario with a large fraction of ionized regions (such as a patchy reionization scenario) and a neutral scenario (e.g., a patchy heating scenario with $\AVXHI\sim 1$). This is one of the major reasons behind the differences in the source parameters posterior distribution compared to the previous case.

Table \ref{tab_mcmc_sourceparamAr} shows the 68 per cent credible intervals on the source parameters disfavoured values at all redshifts for the {\it Single-z} analysis, and at $z=9.1$ for the {\it Joint-z} one. The credible intervals of the {\it Single-z} analysis are similar in trend for all redshifts, except for $\MMINX$ which remains unconstrained at $z=10.1$. The disfavoured credible intervals on $\zeta$ and $\FX$ indicate smaller values towards lower redshifts. The 68 per cent credible interval limits on disfavoured $A_r$ values for the {\it Single-z} analysis correspond to $\TEFF \gtrsim 60$, 55  and 95 K at $z=8.3$, 9.1 and 10.1, respectively, which are equivalent to an excess radio background larger than $140, 100, 207\%$  of the CMB at 1.42 GHz. The {\it Joint-z} analysis posteriors for the source parameters show a trend similar to those from the {\it Single-z} cases, but closer to the $z=9.1$ {\it Single-z} analysis results. The 68 per cent credible intervals on the disfavoured $A_r$ values correspond to $\TEFF \gtrsim 43.2$ K at $z=9.1$ for the {\it Joint-z} analysis, meaning that an excess radio background higher than $57\%$  of the CMB at 1.42 GHz is disfavoured. On the other hand, the 95 per cent credible intervals on the disfavoured $A_r$ value span the entire prior range of $A_r$ for both the {\it Single-z} and {\it Joint-z} cases. 
 
\begin{figure}
\begin{center}
\includegraphics[scale=1.2]{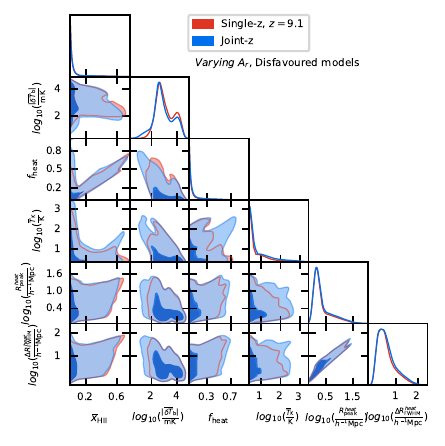}
    \caption{Posterior distribution of the IGM parameters of the models which are disfavoured by the LOFAR upper limits from \citet{lofar2024} at $z= 9.1$ for the {\it Varying} $A_r$ scenario. Red refers to the case when each redshift is considered separately ({\it Single-z}), while blue represents the joint MCMC analysis obtained at $z=$9.1 ({\it Joint-z}). The contour levels in the two-dimensional contour plots refer to the $1\sigma$ and $2\sigma$ credible intervals of the models disfavoured by the LOFAR upper limit at this redshift. The diagonal panels represent the corresponding marginalized probability distributions of each parameter.
    }
   \label{image_igmparam_Ar}
\end{center}
\end{figure}

Figure \ref{image_igmparam_Ar} shows the IGM parameters' posterior distributions for the disfavoured models obtained from the MCMC analysis at $z=9.1$. The 68 (95) per cent credible intervals for the {\it Single-z} analysis case at $z=9.1$ are $\AVXHII \lesssim 0.02 ~(0.46)$,  $\AVTK \lesssim 4.4 ~(44)$ K, $\FHEAT\lesssim 0.05 ~(0.46)$, $1.3\times 10^4 ~(2.6\times 10^4) \gtrsim |\AVTB| \gtrsim 158 ~(91)$ mK, $ \RPEAK\lesssim 3 ~(14) ~\mpcbyh$ and $ \RFWHM\lesssim 5 ~(25)~\mpcbyh$.  The ionization and the thermal states of the disfavoured models are similar to those in the scenario with $A_r=0$, and correspond to extreme models with rare and large ionized/heated regions. We repeat the caveat that the constraints on the IGM parameters might be significantly impacted by the chosen priors on the source parameters (see Appendix \ref{res:appe2} for details).

Table \ref{tab_mcmc_igmAr} shows the disfavoured limits for all three {\it Single-z} and {\it Joint-z} analyses at 68 and 95 per cent credible intervals. In the {\it Single-z} cases, the disfavoured limits on IGM parameters for all redshifts are similar, indicating that the IGM ionization and thermal states producing high amplitudes of the large-scale power spectra are similar. As in the case with $A_r=0$, we also see a consistent redshift evolution of the IGM parameters' disfavoured limits in the {\it Joint-z} analysis case.

\begin{table}
\centering
\caption[]{Constraints on the IGM parameters for the {\it Varying $A_r$} scenario, derived using the LOFAR upper limits from \citet{lofar2024}.
The first column lists the IGM parameters derived from both the {\it Single-z} and {\it Joint-z} analysis, the second refers to the credible intervals of the disfavoured models, while the last three contain the constraints obtained from the analysis at three redshifts. }  
\small
\tabcolsep 3pt
\renewcommand\arraystretch{2.}
   \begin{tabular}{| c | c | c c c c |}
\hline

\makecell{IGM  \\ Parameters \\(MCMC  \\ Model)}  & \makecell{Credible  \\ intervals}  & $z=8.3$ & $z=9.1$ & $z=10.1$ &   \\

\hline
\multirow{2}{4em}{ $\AVXHII$ ({\it Single-z})}
& 68$\%$ & [0, 0.03] & [0, 0.02] & [0, 0.01]  & \\
\cline{2-6}

& 95$\%$ & [0, 0.47] & [0, 0.46] & [0, 0.36] & \\
\cline{1-6}
\multirow{2}{4em}{$\AVXHII$ ({\it Joint-z})}
&  68$\%$ & [0, 0.06] & [0, 0.02] & [0, 0.004]  & \\
\cline{2-6}

& 95$\%$ & [0, 0.8] & [0, 0.43] & [0, 0.16] & \\

\hline

\multirow{2}{4em}{$\FHEAT$ ({\it Single-z})}

& 68$\%$ & [0, 0.07] & [0, 0.05] & [0, 0.03] & \\ 
\cline{2-6}

& 95$\%$ & [0, 0.49] & [0, 0.46] & [0, 0.34] & \\
\cline{1-6}
\multirow{2}{4em}{$\FHEAT$ ({\it Joint-z})}
&  68$\%$ & [0, 0.27] & [0, 0.07] & [0, 0.01] & \\ 
\cline{2-6}

& 95$\%$ & [0, 0.88] & [0, 0.53] & [0, 0.2] & \\
\hline
\multirow{2}{4em}{ $\AVTK$ (K) ({\it Single-z})}

& 68$\%$ & [1.8, 5.0] & [2.2, 4.4] & [2.6, 4.8]  & \\ 
\cline{2-6}

& 95$\%$ & [1.8, 67] & [2.2, 44] & [2.6, 41]  & \\
\cline{1-6}
\multirow{2}{4em}{ $\AVTK$ (K) ({\it Joint-z})}
& 68$\%$ & [1.8, 16] & [2.2, 5.2] & [2.6, 4.5]  & \\ 
\cline{2-6}

& 95$\%$ & [1.8, 977] & [2.2, 101] & [2.6, 186]  & \\

\hline

\multirow{2}{4em}{|$\AVTB$| (mK) ({\it Single-z})}

& 68$\%$ & [174, 11481] & [158,13182] & [396, 11220]  & \\ 
\cline{2-6}

& 95$\%$ & [93, 22908] & [91, 25703] & [144, 29174] & \\
\cline{1-6}
\multirow{2}{4em}{|$\AVTB$| (mK) ({\it Joint-z})}
& 68$\%$ & [25, 3715] & [135,13301] & [177, 16218]  & \\ 
\cline{2-6}

& 95$\%$ & [5, 19952] & [35, 28840] & [158, 28840] & \\

\hline
\multirow{2}{4em}{$\RPEAK$ ($\mpcbyh$) ({\it Single-z})}

& 68$\%$ & [0, 3.0] & [0, 3.0] & [0,3.0]  & \\ 
\cline{2-6}

& 95$\%$ & [0, 14.4] & [0, 14] & [0, 11.2] & \\
\cline{1-6}
\multirow{2}{4em}{$\RPEAK$ ($\mpcbyh$) ({\it Joint-z})}
& 68$\%$ & [0, 6] & [0, 3.0] & [0, 3.0]  & \\ 
\cline{2-6}

& 95$\%$ & [0, 105] & [0, 16.2] & [0, 8.5] & \\

\hline
\multirow{2}{4em}{$\RFWHM$ ($\mpcbyh$) ({\it Single-z})}

& 68$\%$ & [2, 7] & [0, 5] & [0,4]  & \\ 
\cline{2-6}

& 95$\%$ & [0, 27] & [0, 25] & [0, 18] & \\
\cline{1-6}
\multirow{2}{4em}{$\RFWHM$ ($\mpcbyh$) ({\it Joint-z})}
& 68$\%$ & [0, 10] & [0, 6] & [0,4]  & \\ 
\cline{2-6}

& 95$\%$ & [0, 152] & [0,30] & [0, 13.5] & \\

\hline

\end{tabular}
\label{tab_mcmc_igmAr}
\end{table}


\subsection{Analysis including results from other interferometers}
\label{sec:mixed}
Besides LOFAR, several other radio interferometers, such as GMRT, PAPER, MWA and HERA, have produced upper-limits on the 21-cm signal power spectrum in the range $z=8-10$. In this section we consider all available upper limits to derive constraints on the IGM physical properties in this redshift range. Except for LOFAR and MWA, the other upper limits are obtained in a limited small-scale range with $k\gtrsim 0.3 ~\hbympc$. Combining all available upper limits in a MCMC analysis needed to consider a broader $k$ range compared to the earlier cases.

\begin{figure}
\begin{center}
\includegraphics[scale=0.5]{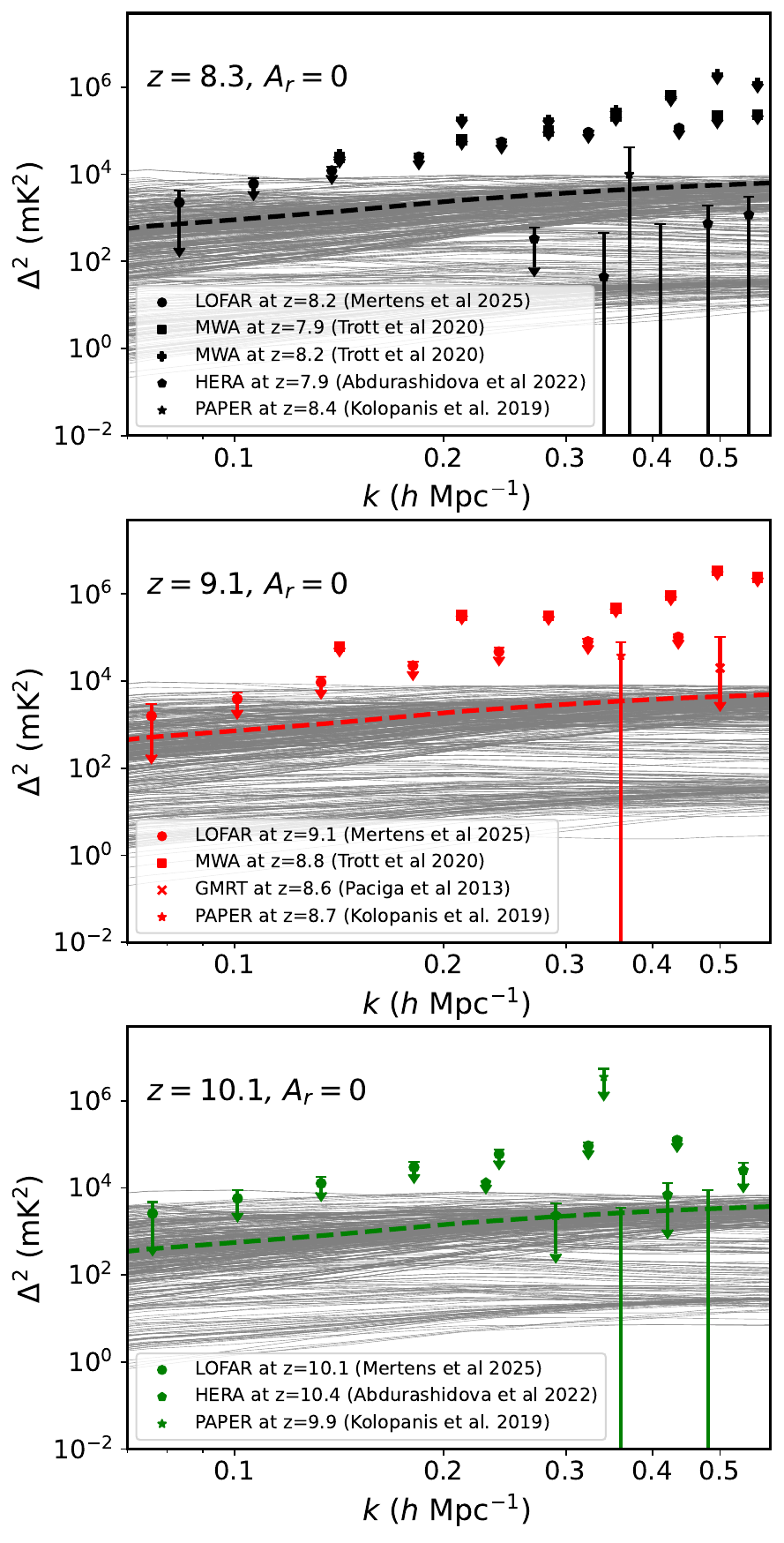}
    \caption{A set of 1000  power spectra randomly chosen out of the $ 10^5$ simulated ones for the scenario with no additional radio background other than the CMB ($A_r = 0$). Panels from top to bottom refer to $z=8.3$, 9.1 and 10.1, respectively. The dashed lines correspond to the power spectrum for a completely neutral IGM with no X-ray heating, and with $\TS=\TK$ at those redshifts.  The down arrow points refer to the different 2$\sigma$ upper limits at those redshifts. }
   \label{image_psts_zallupperlimits}
\end{center}
\end{figure}

\begin{figure}
\begin{center}
\includegraphics[scale=0.75]{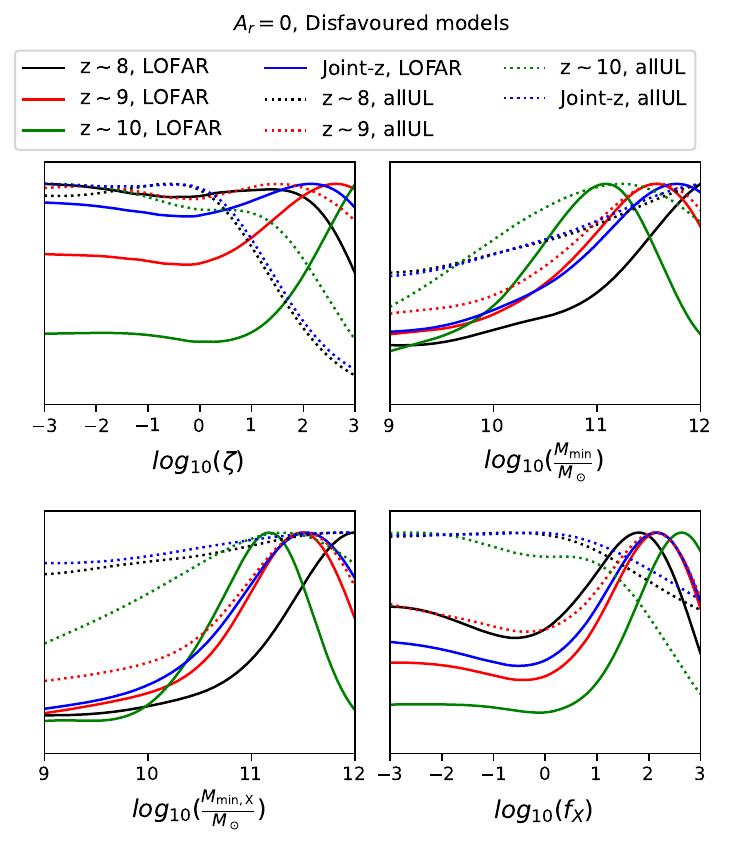}
    \caption{Marginalized probability distribution of the disfavoured models' source parameters at $z\sim8$, 9 and 10 when including upper limits from different radio interferometric observations. The results are for the $A_r=0$ scenario. The black, red and green curves refer to the {\it Single-z} case at $z=8.3$, 9.1 and 10.1, respectively, while the blue curves are for the {\it Joint-z} analysis case. The solid curves represent results derived from the LOFAR upper limits only, while the dotted curves refer to cases where we have used all available upper limits.}
   \label{image_source1dall_Ar0all}
\end{center}
\end{figure}

\begin{figure}
\begin{center}
\includegraphics[scale=1.]{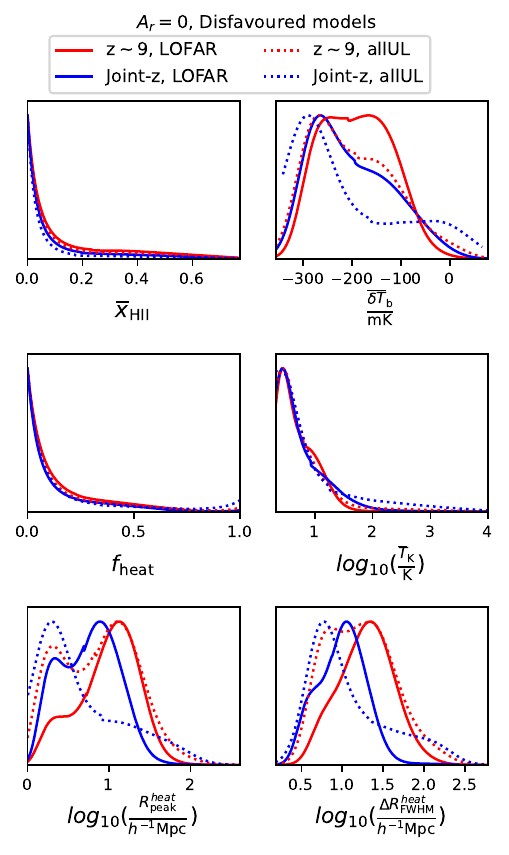}
    \caption{Marginalized probability distribution of the IGM parameters of the disfavoured models at $z\sim9$ when including upper limits from different radio interferometric observations. The results are for the $A_r$ = 0 scenario.  The red and blue curves refer to the {\it Single-z} and {\it Joint-z} cases, respectively. The solid curves represent results derived from the LOFAR upper limits only, while the dotted refer to cases where we have used all available upper limits.}
   \label{image_IGM1dall_Ar0all}
\end{center}
\end{figure}

Figure \ref{image_psts_zallupperlimits} shows those 2$\sigma$ upper limits along with 1000 {\sc grizzly} power spectra randomly chosen from the $10^5$ simulations for $A_r=0$. We note that at $z \sim 8$ only LOFAR \citep{lofar2024} and HERA \citep{Abdurashidova_2023} upper limits can rule out some EoR models at the $2\sigma$ level. In comparison to LOFAR, a significantly larger number of models is ruled out by the HERA best $2\sigma$ upper limit $\DTB(k = 0.34 \hbympc, ~z=7.9, 1128 ~\rm hours) \approx {(21.4)}^{2} ~{\rm mK}^2$, including the scenario of a completely neutral and unheated IGM (dashed line).  This is consistent with the findings of \citet{2022ApJ...924...51A}. The best upper limit from MWA  \citep{2020MNRAS.493.4711T}, $\DTB(k=0.14 \hbympc,~z=7.8, 51 \rm ~hours)\approx (154)^2 ~{\rm mK}^2$, is still above the extreme EoR models power spectra by a factor of $\approx 2$. The best $2\sigma$ result from PAPER is $\DTB(k=0.37 \hbympc,~z=8.4, 135 \rm ~nights)\approx (205)^2 ~{\rm mK}^2$, which is larger than the predicted power spectra by a factor of $\gtrsim 4$.

At $z\sim9$, the LOFAR upper limits from \citet{lofar2024} are the best available results, and remain the only ones which rule out a set of extreme EoR scenarios. The best $2\sigma$ upper limit from GMRT \citep{paciga13} and MWA \citep{2020MNRAS.493.4711T} at this redshift are $\DTB(k=0.5 \hbympc,~z=8.6, 50 \rm ~hours)\approx (248)^2 ~{\rm mK}^2$ and $\DTB(k=0.14 \hbympc,~z=8.8, 51 \rm ~hours)\approx (250)^2 ~{\rm mK}^2$, respectively, i.e. more than $\approx6$ times larger than our extreme models. 
The best $2\sigma$ result from PAPER is $\DTB(k=0.36 \hbympc,~z=8.7, 135 \rm ~nights)\approx (281)^2 ~{\rm mK}^2$, which is larger than the {\sc grizzly} power spectra by a factor of at least $\approx8$. We note that the neutral and unheated IGM scenario at $z\sim9$ (dashed curve) hasn't been ruled out by any of these upper limits.

Similarly to $z\sim8$, also at $z\sim10$ only LOFAR and HERA upper limits can rule out some extreme EoR models, as shown in the bottom panel of Fig. \ref{image_psts_zallupperlimits}. The best $2\sigma$ upper limit result from HERA at this redshift is $\DTB(k = 0.36 \hbympc, ~z=10.4, 1128 ~\rm hours) \approx {(59)}^{2} ~{\rm mK}^2$, which is marginally able to rule out the neutral and unheated EoR scenario. Compared to $z\sim8$, the number of models ruled out by HERA  at $z\sim10$ is smaller. The only other interferometer which has collected data at this redshift is PAPER, with a best $2\sigma$ result of $\DTB(k=0.34 \hbympc,~z=9.9, 135 \rm ~nights)\approx (1923)^2 ~{\rm mK}^2$, which is several orders of magnitude higher than our theoretical predictions.

Next we perform an MCMC analysis on the disfavoured models for the $A_r=0$ scenario, including all available upper limit results at $z\sim8$, 9 and 10. In this case, we have included one $k-$bin corresponding to the best upper limit results from each telescope in addition to the three LOFAR $k-$bins considered in our previous analysis. We denote this as the {\it allUL} scenario. The 1D posterior distribution of the disfavoured source parameters for {\it allUL} and $A_r=0$ is shown (dotted lines) in Fig. \ref{image_source1dall_Ar0all} along with the LOFAR only case (solid). The 68 per cent disfavoured credible intervals from the {\it Single-z} analysis at $z\sim9$ for the {\it allUL} case are $\MMIN \gtrsim 2.4\times 10^{10} ~\MSUN$ and $\MMINX \gtrsim 3.1\times 10^{10} ~\MSUN$, while the limits for $\zeta$ and  $f_X$ span the entire parameter ranges. The 95 per cent disfavoured credible intervals span the entire parameter ranges for both the {\it Single-z} and {\it Joint-z} analysis. For the {\it Joint-z} and  $\it allUL$ analysis case, the 68 per cent disfavoured credible intervals at $z\sim 9$ are $\zeta\lesssim 1.8$ and $\MMIN \gtrsim 1.4\times 10^{10} ~\MSUN$, while the limits for $\MMINX$ and  $f_X$ span the entire parameter ranges, similarly to the {\it Single-z} analysis. 
The presence of small-scale power spectra upper limits in the MCMC analysis, causes a significant change in the marginalized probability distribution of the source parameters with respect to the LOFAR-only case. The minimum change is observed at $z\sim 9$, as the LOFAR upper limits at the large-scales are the best ones and thus dominate the MCMC analysis. Instead, at the other two redshifts the posteriors are mainly determined by the HERA upper limits.

\begin{figure}
\begin{center}
\includegraphics[scale=1.]{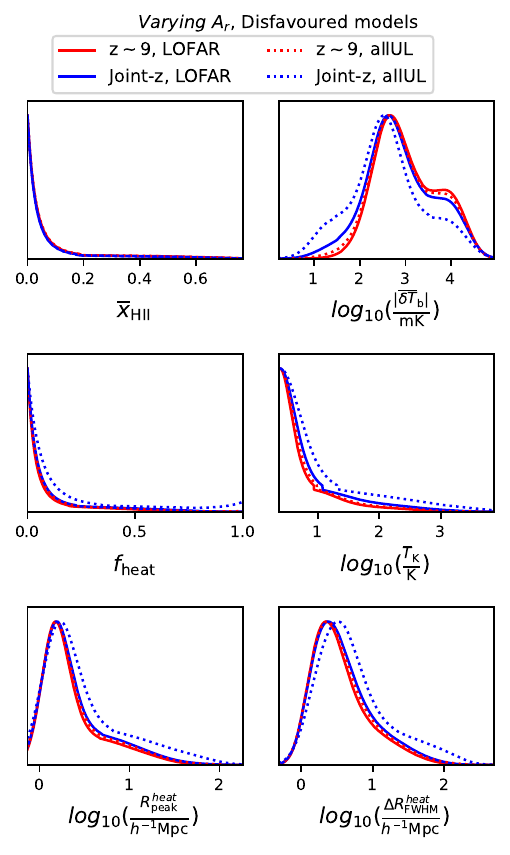}
    \caption{Same as Fig. \ref{image_IGM1dall_Ar0all} but for the $Varying ~A_r$ scenario.}
   \label{image_IGM1dall_Arall}
\end{center}
\end{figure}

In Fig. \ref{image_IGM1dall_Ar0all} we present the 1D marginalized probability distribution of the disfavoured models' IGM parameters at $z\sim 9$ for the {\it allUL} scenario with $A_r=0$, which in the {\it Single-z} analysis case have $\AVXHII \lesssim 0.08 ~(0.54)$,  $\AVTK \lesssim 7 ~(38)$ K, $-308 ~(-344) \lesssim \AVTB \lesssim -134 ~(27)$  mK, $ \RPEAK \lesssim 21 ~(44) ~\mpcbyh$ and $ \RFWHM\lesssim 35 ~(104)~\mpcbyh$ at 68 (95) per cent credible intervals, while $\FHEAT$ spans the entire range. With the {\it Joint-z} analysis we instead obtain $\AVXHII \lesssim 0.02 ~(0.42)$,  $\AVTK \lesssim 6 ~(270)$ K, $-307 ~(-307) \lesssim \AVTB \lesssim -176 ~(13)$  mK, $\FHEAT \lesssim 0.12 (X)$, $ \RPEAK \lesssim 7 ~(63) ~\mpcbyh$ and $ \RFWHM\lesssim 20 ~(151)~\mpcbyh$. While the constraints on $\AVXHII, ~\FHEAT, ~\AVTK$ look similar for the {\it allUL} and LOFAR-only cases, significant differences are seen for the other parameters. Indeed, due to the presence of the small-scale upper limits with a large error, more reionization scenarios with smaller emission/ionized regions are disfavoured in the {\it allUL} scenario.

Finally, Fig. \ref{image_IGM1dall_Arall} presents the 1D marginalized probability distribution of the disfavoured IGM parameters at $z\sim 9$ for the {\it allUL} and {\it Varying $A_r$} scenario. The 68 (95) per cent credible intervals of the disfavoured models for the {\it Single-z} analysis are $\AVXHII \lesssim 0.02 ~(0.46)$, $\FHEAT \lesssim 0.06~(0.5)$ $\AVTK \lesssim 5 ~(55)$ K, $ 131~(62) \lesssim |\AVTB| \lesssim 3812 ~(28184)$  mK, $ \RPEAK \lesssim 5 ~(17) ~\mpcbyh$ and $ \RFWHM\lesssim 7 ~(32)~\mpcbyh$. For the {\it Joint-z} analysis, these intervals instead are $\AVXHII \lesssim 0.02 ~(0.42)$,  $\AVTK \lesssim 6 ~(371)$ K, $51 ~(13) \lesssim |\AVTB| \lesssim 2238 ~(19952)$  mK, $\FHEAT \lesssim 0.12 (X)$, $ \RPEAK \lesssim 5 ~(35) ~\mpcbyh$ and $ \RFWHM\lesssim 9 ~(76)~\mpcbyh$. Table \ref{tab_mcmc_igmArallUL} presents the 68 and 95 per cent credible intervals  at all three redshifts for both the {\it Single-z} and {\it Joint-z} analysis. Contrary to Fig. \ref{image_IGM1dall_Ar0all}, here we see that the constraints on all our IGM parameters look similar to the {\it allUL} and LOFAR-only case in the {\it Single-z} analysis. The difference on the posterior distributions for the {\it Joint-z} case as shown in Fig. \ref{image_IGM1dall_Arall} is also not as significant as found in Fig. \ref{image_IGM1dall_Ar0all}. While the LOFAR $z=9.1$ upper limits, being the best ones at this redshift, try to keep the MCMC results similar for the {\it allUL} and LOFAR-only {\it Single-z} case, the effect of source parameters' priors in the {\it Varying $A_r$} scenario also play an important role in producing similar posterior distributions.

We have also performed an MCMC analysis for the other two redshifts, the results of which can be found in Appendix \ref{appen:all}. This includes the constraints on the IGM parameters for the {\it allUL} case in Tables \ref{tab_mcmc_igmAr0allUL} and \ref{tab_mcmc_igmArallUL}. Figure \ref{image_IGM1dall_Ar0z8all} and \ref{image_IGM1dall_Ar0z10all} show the comparison between the {\it allUL} and LOFAR-only $A_r=0$ scenario for redshifts $\sim8$ and 10, respectively. Clearly, the 1D posterior distributions are biased by the HERA upper limits, making the {\it allUL} and LOFAR-only PDFs different for most of the IGM parameters. This is also true for Fig. \ref{image_IGM1dall_Arz8all} and \ref{image_IGM1dall_Arz10all}, which show the same comparison but for the {\it Varying $A_r$} scenario for redshifts $\sim8$ and 10, respectively.

\section{Summary \& Discussion}
\label{sec:con}

Recently, \citet{lofar2024} analysed 140 hours of LOFAR data for redshifts 8.3, 9.1 and 10.1, and produced new upper limits on the dimensionless spherically averaged power spectrum of the 21-cm signal (see Table \ref{tab_obs}). Here we have used those upper limits to infer the physical state of the IGM, as well as the properties of the ionizing sources, at those redshifts. As the upper limit results are still weak, we have derived constraints on the disfavoured rather than favoured models of reionization. 

Our framework uses a databse of hundreds of thousands of reionization models simulated with the {\sc grizzly} code, which employs a one-dimensional radiative transfer approach.  The models are obtained by changing the following parameters associated with source properties:  ionization efficiency ($\zeta$), minimum mass of the UV emitting halos ($\MMIN$), minimum mass of X-ray emitting halos ($\MMINX$), X-ray heating efficiency ($f_X$) and the amplitude of a radio background in excess of the CMB ($A_r$). In addition to the power spectra of the 21-cm signal, various other quantities are derived from the reionization models to characterise the ionization and thermal state of the IGM, such as volume averaged ionization fraction ($\AVXHII$), volume averaged gas temperature of the partially ionized IGM ($\AVTK$), mass averaged brightness temperature ($\AVTB$), volume fraction of the heated region ($\FHEAT$), size of the heated regions at which the PDF of the sizes peaks ($\RPEAK$), and FWHM of the PDFs ($\RFWHM$). We then use an MCMC framework to derive constraints on the source parameters and IGM-related quantities. We consider two different scenarios, one in which there is no excess radio background, and one in which there is one, characterized by a parameter $A_r$. We perform one MCMC analysis considering each redshift independently (referred to as the {\it Single-z} analysis case), and another one in which we derived a joint likelihood for the three redshifts ({\it Joint-z} analysis). The constraints obtained on the disfavoured source and IGM parameters are based on the chosen uniform priors of the source parameters as presented in Table \ref{tab_source_param}. We highlight that our source parameters are assumed to be redshift independent, and thus the results of the {\it Joint-z} analysis should be interpreted with caution.

The main findings of the paper can be summarized as follows. 

\begin{itemize}
\item More than 10 per cent of the simulation database with $A_r=0$ are well above $\DTBLOFARMEAN (k, z) - \DTBLOFARSIGMA (k, z)$ in at least one $k-$bin at all three redshifts, indicating that these models have a high probability of being ruled out.  

\item The models disfavoured by both the {\it Single-z} and the {\it Joint-z} analysis for the scenario with $A_r=0$ typically have a small number density of bright UV and X-ray emitting sources at all redshifts. More specifically, in the {\it Single-z} analysis, the 68 per cent credible intervals on the disfavoured $\MMIN$ and $\MMINX$ values at $z=9.1$ are $\gtrsim 3 \times 10^{10} ~\MSUN$ and $\gtrsim 6 \times 10^{10} ~\MSUN$, respectively, while $\zeta$ and $\FX$ remain unconstrained (see Table \ref{tab_mcmc_sourceparamAr0} for the constraints at all redshifts). We find that the 95 per cent disfavoured credible intervals of the source parameters span the entire prior ranges. The MCMC analysis results in the {\it Joint-z} case are significantly biased by the upper limits at $z=9.1$, which are the strongest among the three redshifts considered here. 

\item For $A_r=0$, the disfavoured limits on the IGM parameters at $z=9.1$ are $\AVXHII \lesssim 0.13 ~(0.55)$,  $\AVTK \lesssim 7.3 ~(21)$ K, $\FHEAT \lesssim 0.18 ~(0.57)$, $-276 ~(-317) \lesssim \AVTB \lesssim -123 ~(-61)$  mK, $ \RPEAK \lesssim 28 ~(40) ~\mpcbyh$ and $ \RFWHM\lesssim 46 ~(81)~\mpcbyh$ at 68 (95) per cent credible interval level (see Table \ref{tab_mcmc_igmAr0} for details), indicating extreme reionization models with rare and large ionized/heated regions. The {\it Single-z} analysis gives similar credible intervals at all redshifts, showing that only a few IGM physical states can produce very high amplitudes of the large-scale power spectrum. As for the source parameters, the results of the {\it Joint-z} analysis for the IGM parameters are mainly biased by the $z=9.1$ upper limits. They show a consistent redshift evolution, e.g. the 95 per cent credible interval limits on $\AVXHII$ are $\lesssim 0.85$, $\lesssim 0.46$ and $\lesssim 0.17$ at $z=8.3$, 9.1 and 10.1, respectively.

\item Also when considering an excess radio background, the models disfavoured by both the {\it Single-z} and the {\it Joint-z} analysis have a low number density of bright UV and X-ray emitting sources at all redshifts. More specifically, in the {\it Single-z} analysis the 68 per cent credible intervals on the source parameters at $z=9.1$ are $\zeta \lesssim 2.4$, $\MMIN \gtrsim 1.7\times 10^{10} ~\MSUN$, $\MMINX \gtrsim 1.6\times 10^{10} ~\MSUN$, $f_X \lesssim 5.2$ and $A_r \gtrsim 4.6$  (see Table \ref{tab_mcmc_sourceparamAr} for the constraints at all redshifts). We find that the 95 per cent disfavoured credible intervals span the entire prior ranges of the source parameters for both the  {\it Single-z} and {\it Joint-z} analysis. We also find that the 68 per cent credible intervals are  $A_r \gtrsim 7.9$ and 7.2  at $z=8.3$ and 10.1, respectively, suggesting that an excess radio background which is at least $140, 100,$ and $207$ per cent of the CMB at 1.42 GHz is disfavoured at $z=8.3$, 9.1 and 10.1, respectively. As for $A_r=0$, the {\it Joint-z} analysis is biased by the upper limits at $z=9.1$. In terms of the excess radio background, in this case we find that the 68 per cent limits on the disfavoured models' $A_r$ at $z=9.1$ correspond to an excess radio background which is $\gtrsim 57\%$  of the CMB at 1.42 GHz. 

\item For such a radio background model, the disfavoured limits on the IGM parameters are $\AVXHII \lesssim 0.02 ~(0.46)$,  $\AVTK \lesssim 4.4 ~(44)$ K, $\FHEAT\lesssim 0.05 ~(0.46)$, $1.3\times 10^4 ~(2.6\times 10^4) \gtrsim |\AVTB| \gtrsim 158 ~(91)$ mK, $ \RPEAK\lesssim 3 ~(14) ~\mpcbyh$ and $ \RFWHM\lesssim 5 ~(25)~\mpcbyh$ at $z=9.1$ for the {\it Single-z} case at 68 (95) per cent credible interval level (see Table \ref{tab_mcmc_igmAr} for all redshifts' constraints). This indicates that the disfavoured ionization and thermal states are similar to those in the scenario with $A_r=0$, and correspond to extreme reionization models with rare and large ionized/heated regions. Similarly to the $A_r=0$ scenario, we also find a consistent redshift evolution of the IGM parameters' disfavoured limits in the {\it Joint-z} analysis for the {\it Varying} $A_r$ case.

\item The inclusion of upper limits from other radio interferometric observations ({\it allUL} case) in our study significantly increases the disfavoured parameter space of both the source and IGM parameters. We find that the HERA upper limits dominate the Bayesian analysis at $z\sim8$ and 10, while LOFAR ones prevail at $z\sim9$. The HERA upper limits rule out a neutral and unheated IGM at $z\sim8$ at a $2\sigma$ level. We find that the constraints on $\AVXHII, ~\FHEAT, ~\AVTK$ are similar for the {\it allUL} and LOFAR-only cases, while those on the other parameters are significantly different.

\end{itemize}

The constraints derived in this paper improve on those obtained in \citet{2020MNRAS.493.4728G} using the earlier LOFAR upper limits at $z=9.1$ from \citet{2020MNRAS.493.1662M}. Indeed, in the absence of a radio background in excess of the CMB, the 95 per cent credible intervals on the disfavoured models' IGM parameters obtained in  \citet{2020MNRAS.493.4728G} were $\FHEAT \lesssim 0.34$, $\AVTK\lesssim 160$ K, $\RPEAK \lesssim 70 ~\mpcbyh$ and $\RFWHM \lesssim 110 ~\mpcbyh$, while we found two regimes of disfavoured credible intervals for $\AVXHII$, i.e. $\AVXHII\lesssim0.08$ and $0.45\lesssim \AVXHII \lesssim 0.62$. We note that there are some differences between the analysis done here and in \citet{2020MNRAS.493.4728G}. For example, the latter fixed $\MMIN=10^9~\MSUN$ while varying $\zeta$, $\MMINX$ and $\FX$ with different prior ranges, and they did not include redshift space distortions in their simulated power spectra, resulting in under-predictions of the large-scale power spectrum for several scenarios. The improvement in the present version of our interpretation framework, as well as in the observed upper limits, has enabled us to exclude more reionization models and to produce better posterior distributions for both the disfavoured source and IGM parameters.

Previously. \citet{2020MNRAS.498.4178M} constrained the excess radio background as $0.1-9.6$ per cent of the CMB at 1.42 GHz using the LOFAR upper limits at $z=9.1$ from \citet{2020MNRAS.493.1662M}. We note that a direct comparison to these constraints is not possible, as they were obtained by marginalizing models which are allowed under those upper limits, while our results are derived from the marginalization of the excluded models. In  \citet{2021MNRAS.503.4551G}, an excess radio background of at least 25 per cent of the CMB at 1.42 GHz was excluded using MWA upper limits in the range $z=6.5-8.7$ from \citet{2020MNRAS.493.4711T} with 68 per cent credible intervals on $A_r$.

The MCMC results presented in this study are dependent on the prior ranges of the source parameters. This is true not only for disfavoured models, but also for allowed ones. Since the volume in parameter space is set by the priors, their impact can be very large, in particular if the interested scale is unknown and a flat prior in log space is employed. Therefore, the results strongly depend on the lower and upper limits set on the model parameters. In this work, we find that the constraints on $\AVXHII$, $\FHEAT$ and $\AVTK$ are significantly affected by the chosen priors on the source parameters (see Appendix \ref{res:appe2}). To improve on this, observational data on high-$z$ galactic properties,  such as the luminosity function from the James Webb Space Telescope \citep[JWST; e.g.,][]{Donnan_2022, Castellano_2023}, can be used to increase our understanding of the source parameters prior ranges, and thus to improve the constraints on the source and IGM parameters of the excluded models.

The upper limits on the 21-cm signal power spectrum used in this study contain excess noise and thus are conservative in nature. This is also applicable to the previous LOFAR upper limits from \cite{2017ApJ...838...65P}, \cite{2020MNRAS.493.1662M}, and \cite{emilio2024}. The source of this excess noise is still uncertain \citep[see e.g.,][]{Gan2022}. Recently,  \citet{2024MNRAS.534L..30A} considered applying a bias correction for excess noise as well and, in this case, reported a 2-$\sigma$ upper limit of $\DTB(k = 0.075 \hbympc, ~z=9.1) \approx (25)^2 ~{\rm mK}^2$ based on 141 hours of LOFAR data used in \cite{2020MNRAS.493.1662M}. In Appendix \ref{res:appe1} we perform an inference study on the \citet{2024MNRAS.534L..30A} upper limits, finding that the 95 per cent credible intervals of the IGM parameters of the allowed models are $\AVXHII \lesssim 0.75$,  $27 \lesssim \AVTK \lesssim 7\times10^3$ K, $\FHEAT \gtrsim 0.62$, $4 \lesssim \AVTB \lesssim 44$ mK, $ 3 \lesssim \RPEAK\lesssim 160 ~\mpcbyh$ and $ 4 \lesssim \RFWHM\lesssim 316~\mpcbyh$ for the {\it Varying} $A_r$ {\it Single-z} scenario at $z=9.1$. As in this case the upper limits are stronger, the results are on the allowed rather than the excluded models,  and thus cannot be directly compared to the results presented in the rest of the paper. We also caution the reader that the robustness of the \citet{2024MNRAS.534L..30A} upper limits needs further investigation, as also discussed in the original paper. For this reason, here we have focused our attention on the constraints obtained from the weaker upper limits of \citet{lofar2024}.

We also note that, in principle, an inference study such as the one discussed here should be included in the data analysis pipeline, together with the foreground mitigation, rather than as a separate step. This, though, is beyond the scope of this paper, but will be considered in future studies.

It should be realised that the current LOFAR upper limits on $\DTB$ are rather large, and thus the constraints obtained on the disfavoured IGM and source parameters are not yet very strong. However, this study illustrates the potential of the 21-cm signal observation as a probe of the EoR, as even before an actual detection of the signal becomes available, more stringent upper limits on $\DTB$ will provide stronger constraints on the IGM parameters during the first billion years of our Universe. Additionally, other probes of this era such as the CMB, the global 21-cm signal, and high-$z$ sources (e.g. galaxies, quasars and fast radio bursts) will strengthen the understanding of the IGM during this crucial period in the Universe's history. In particular, the detection of many faint galaxies at $z \gtrsim 6$ with JWST, complemented by observations with the Atacama Large Millimetre Array and the future European Extremely Large Telescope, will improve our understanding of the properties of early galaxies. This information will be fundamental to improve the prediction of the 21-cm signal by calibrating the source model with these observations (see e.g. the {\sc grizzly} extension {\sc polar}; \citealt{2023MNRAS.522.3284M}), and thus providing us with a much deeper understanding of the EoR sources and IGM properties.

\begin{acknowledgements}
RG acknowledges support from SERB, DST Ramanujan Fellowship no. RJF/2022/000141. We acknowledge that the N-body simulation used in this paper has been achieved using the PRACE Research Infrastructure resources Curie based at the Tr$\grave{\rm e}$s Grand Centre de Calcul (TGCC) operated by CEA near Paris, France and Marenostrum based in the Barcelona Supercomputing Center, Spain (PRACE under PRACE4LOFAR grants 2012061089 and 2014102339 as well as under the Multi-scale Reionization grants 2014102281 and 2015122822). SKG is supported by NWO grant number OCENW.M.22.307. LVEK, SAB, KC, SG, CH and SM acknowledge the financial support from the European Research Council (ERC) under the European Union’s Horizon 2020 research and innovation programme (Grant agreement No. 884760, ``CoDEX''). FGM acknowledges the support of the PSL Fellowship. EC acknowledge support from the Centre for Data Science and Systems Complexity (DSSC), Faculty of Science and Engineering at the University of Groningen, and the Ministry of Universities and Research (MUR) through the PRIN project `Optimal inference from radio images of the epoch of reionization'. SZ acknowledge support by the Israel Science Foundation (grant no. 1388/24). GM’s research is supported by the Swedish Research Council project grant 2020-04691\_VR. MC's research is supported by the Center for Fundamental Physics of the Universe(CFPU) Postdoctoral fellowship, Brown University, Providence, RI.

\end{acknowledgements}

\bibliographystyle{aa}
\bibliography{mybib}

\begin{appendix}
\section{Effect of the source parameter priors on the IGM parameters constraints }
\label{res:appe2}

\begin{figure}
\begin{center}
\includegraphics[scale=1.]{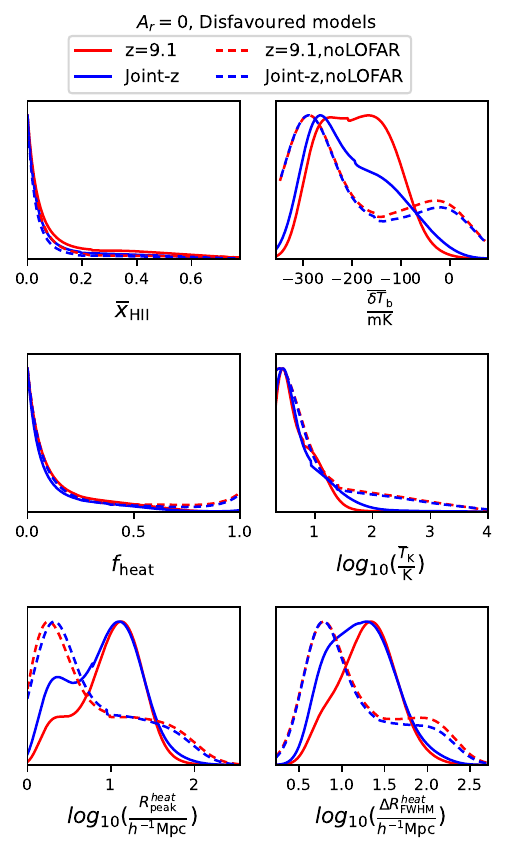}
    \caption{Marginalized probability distribution of the IGM parameters of the disfavoured models for a constant exclusion likelihood. The red and blue curves refer to the {\it Single-z} and {\it Joint-z} cases at  $z=9.1$, respectively. The solid curves represent results from the main text, while the dashed refers to cases where we have used $\mathcal{L}_{\mathrm{ex, single}-z}(\bm{\theta}, z=9.1)=0.5$ and $\mathcal{L}_{\mathrm{ex, joint}-z}(\bm{\theta}, z=9.1)=0.5$.}
   \label{image_IGM1dall_Ar0}
\end{center}
\end{figure}

In the study described in the main text we use source parameters as input for the MCMC analysis. After this the list of source parameter values is used to produce a corresponding list of IGM parameters, which is then employed to constrain the IGM parameters themselves. This method might introduce non-uniform posterior distributions of the IGM parameters even when the LOFAR upper limits are not considered in the analysis. In this section, we aim to investigate the effect of the source parameters' priors on the constraints on the IGM parameters, by adopting the following approach.  We fix to 0.5 the exclusion likelihood shown in Eqs. (\ref{equ:like}) and (\ref{equ:likejoint}), independently from the LOFAR upper limits, but still use the priors on the maximum value of $\AVXHII$, and we denote this as the {\it no-LOFAR} scenario. In this case, the posterior distribution of the source parameters is expected to be independent of the choice of the constant likelihood, while the posterior distribution of the disfavoured IGM parameter values is entirely determined by the chosen priors on the source parameters. 

We consider the four source parameters  for the $A_r=0$ scenario from Sect.~\ref{sec:noRBG}, but with a constant likelihood. The dashed curves in Fig. \ref{image_IGM1dall_Ar0} refer to the resulting marginalized probability distribution of the IGM parameters of the disfavoured models, for both the {\it Single-z} and {\it Joint-z} analysis. The distribution shows the effect of the source priors on the posteriors for the IGM parameters. 
The disfavoured models have $\AVXHII \lesssim 0.024 ~(0.56)$,  $\AVTK \lesssim 7.2 ~(618)$ K, no constraints on $\FHEAT$, $-315 \lesssim \AVTB \lesssim -129 ~(18.1)$  mK, $ \RPEAK \lesssim 10.8 ~(83.1) ~\mpcbyh$ and $ \RFWHM\lesssim 87 ~(207)~\mpcbyh$ at 68 (95) per cent credible intervals for the {\it Single-z} {\it no-LOFAR} analysis case. For the {\it Joint-z} {\it no-LOFAR} analysis case, the disfavoured models' 68 (95) per cent credible intervals are $\AVXHII \lesssim 0.019 ~(0.43)$,  $\AVTK \lesssim 6 ~(562)$ K, no constraints on $\FHEAT$, $-315 \lesssim \AVTB \lesssim -149 ~(18.1)$  mK, $ \RPEAK \lesssim 7 ~(36) ~\mpcbyh$ and $ \RFWHM\lesssim 23 ~(93)~\mpcbyh$. For comparison, the solid curves show the probability distribution of the IGM parameters of the disfavoured models obtained in Section~\ref{sec:noRBG}. Any differences between the solid and dashed curves are due to the effect of using the LOFAR upper limits. Such differences are clearly visible for $\AVTB$, $\RPEAK$ and $\RFWHM$, they are less prominent for $\AVXHII$, $\FHEAT$ and $\AVTK$. This suggests that the LOFAR upper limits provide very weak constraints on the latter three IGM parameters. The fact that the posterior for $\AVTB$ is affected by the LOFAR upper limits but the other averages ($\AVXHII$, $\FHEAT$ and $\AVTK$) are not, suggests that the correlations between $\delta$, $\XHII$ and $\TK$ play important roles in determining the 21-cm signal power spectra \citep[see e.g.,][]{2024PhRvD.110b3543S}.

\section{Inference including all available upper limit results at $z\sim 8$ and 10}
\label{appen:all}

While the results from the MCMC analysis which includes all available upper limits ($allUL$ case) at $z\sim9$ have been discussed in Sect.~\ref{sec:mixed}, here we present the results at $z\sim8$ and 10. The 68 per cent disfavoured credible intervals of the source parameters for the $A_r=0$ and {\it Single-z} scenario at $z\sim8$ are $\zeta\lesssim 1.6$ and $\MMIN \gtrsim 1.4\times 10^{10} ~\MSUN$, while the limits for $\MMINX$ and  $f_X$ span the entire parameter ranges (see Fig. \ref{image_source1dall_Ar0all}). The same intervals for $z\sim 10$ are $\zeta\lesssim 5$,$\MMIN \gtrsim 1.7\times 10^{10} ~\MSUN$, $\MMINX \gtrsim 1.6\times 10^{10} ~\MSUN$, and $f_X\lesssim 4$.  In this case, the source parameters' 95 per cent disfavoured credible intervals span the entire parameter ranges for both the redshifts. 
The 1D marginalized probability distribution of the source parameters at $z\sim 8$ and 10 for the $\it allUL$ case are significantly different from the LOFAR-only case, as HERA upper limits dominate the MCMC analysis at these two redshifts.

\begin{table}
\centering
\caption[]{Constraints on the IGM parameters for the $A_r=0$ and {\it allUL} scenario for both the {\it Single-z} and {\it Joint-z} analysis at three redshifts.}
\small
\tabcolsep 3pt
\renewcommand\arraystretch{2.}
   \begin{tabular}{| c | c | c c c c |}
\hline

\makecell{IGM  \\ Parameters \\(MCMC  \\ Model)}  & \makecell{Credible  \\ intervals}  & $z=8.3$ & $z=9.1$ & $z=10.1$ &   \\

\hline
\multirow{2}{4em}{ $\AVXHII$ ({\it Single-z})}
& 68$\%$ & [0, 0.05] & [0, 0.08] & [0, 0.01]  & \\
\cline{2-6}

& 95$\%$ & [0, 0.78] & [0, 0.54] & [0, 0.35] & \\
\cline{1-6}
\multirow{2}{4em}{$\AVXHII$ ({\it Joint-z})}
&  68$\%$ & [0, 0.06] & [0, 0.02] & [0, 0.01]  & \\
\cline{2-6}

& 95$\%$ & [0, 0.79] & [0, 0.42] & [0, 0.15] & \\

\hline

\multirow{2}{4em}{$\FHEAT$ ({\it Single-z})}

& 68$\%$ & [X] & [X] & [0, 0.03] & \\ 
\cline{2-6}

& 95$\%$ & [X] & [X] & [0, 0.36] & \\
\cline{1-6}
\multirow{2}{4em}{$\FHEAT$ ({\it Joint-z})}
&  68$\%$ & [X] & [0, 0.12] & [0, 0.03] & \\ 
\cline{2-6}

& 95$\%$ & [X] & [X] & [X] & \\
\hline
\multirow{2}{4em}{ $\AVTK$ (K) ({\it Single-z})}

& 68$\%$ & [1.8, 10] & [2.2, 7] & [2.6, 5]  & \\ 
\cline{2-6}

& 95$\%$ & [1.8, 2180] & [2.2, 38] & [2.6, 12]  & \\
\cline{1-6}
\multirow{2}{4em}{ $\AVTK$ (K) ({\it Joint-z})}
& 68$\%$ & [1.8, 13] & [2.2, 6] & [2.6, 5]  & \\ 
\cline{2-6}

& 95$\%$ & [1.8, 1737] & [2.2, 270] & [2.6, 79]  & \\

\hline

\multirow{2}{4em}{$\AVTB$ (mK) ({\it Single-z})}

& 68$\%$ & [-352, 3.2] & [-308,-134] & [-285, -241]  & \\ 
\cline{2-6}

& 95$\%$ & [-375, 42] & [-344, 27] & [-285, -117] & \\
\cline{1-6}
\multirow{2}{4em}{$\AVTB$ (mK) ({\it Joint-z})}
& 68$\%$ & [-349, 7] &[-307,-176] & [-285, -239]   & \\ 
\cline{2-6}

& 95$\%$ &  [-374, 45] & [-307, 13] & [-285, -15] & \\

\hline
\multirow{2}{4em}{$\RPEAK$ ($\mpcbyh$) ({\it Single-z})}

& 68$\%$ & [0, 17] & [0, 21] & [0, 7]  & \\ 
\cline{2-6}

& 95$\%$ & [0, 169] & [0, 44] & [0, 17] & \\
\cline{1-6}
\multirow{2}{4em}{$\RPEAK$ ($\mpcbyh$) ({\it Joint-z})}
& 68$\%$ & [0, 21] & [0, 7] & [0, 5]  & \\ 
\cline{2-6}

& 95$\%$ & [0, 169] & [0, 63] & [0, 66] & \\

\hline
\multirow{2}{4em}{$\RFWHM$ ($\mpcbyh$) ({\it Single-z})}

& 68$\%$ & [6, 112] & [6, 35] & [2,12]  & \\ 
\cline{2-6}

& 95$\%$ & [3, 316] & [3, 104] & [3, 36] & \\
\cline{1-6}
\multirow{2}{4em}{$\RFWHM$ ($\mpcbyh$) ({\it Joint-z})}
& 68$\%$ & [0, 21] & [8, 20]  & [0, 11]  & \\ 
\cline{2-6}

& 95$\%$ & [0, 295] & [3, 151] & [0, 66] & \\

\hline

\end{tabular}
\label{tab_mcmc_igmAr0allUL}
\end{table}

\begin{figure}
\begin{center}
\includegraphics[scale=1.]{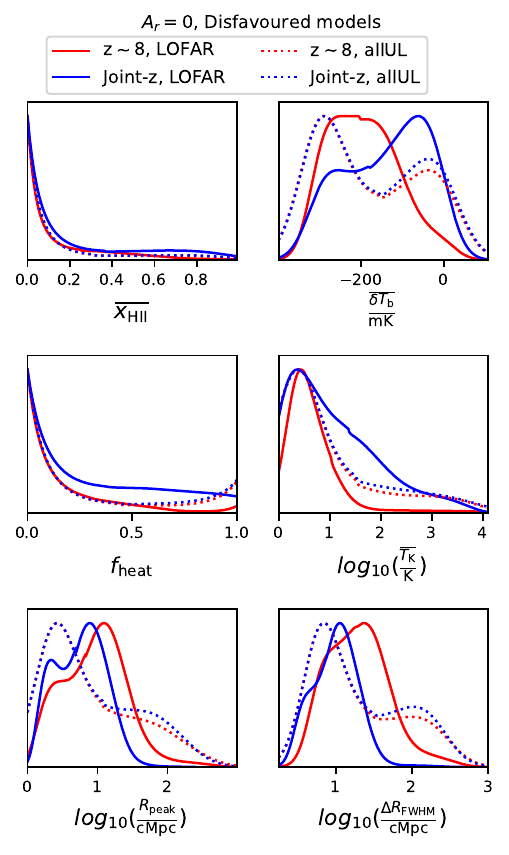}
    \caption{Same as Fig. \ref{image_IGM1dall_Ar0z10all} but for $z\sim8$.}
   \label{image_IGM1dall_Ar0z8all}
\end{center}
\end{figure}

\begin{figure}
\begin{center}
\includegraphics[scale=1.]{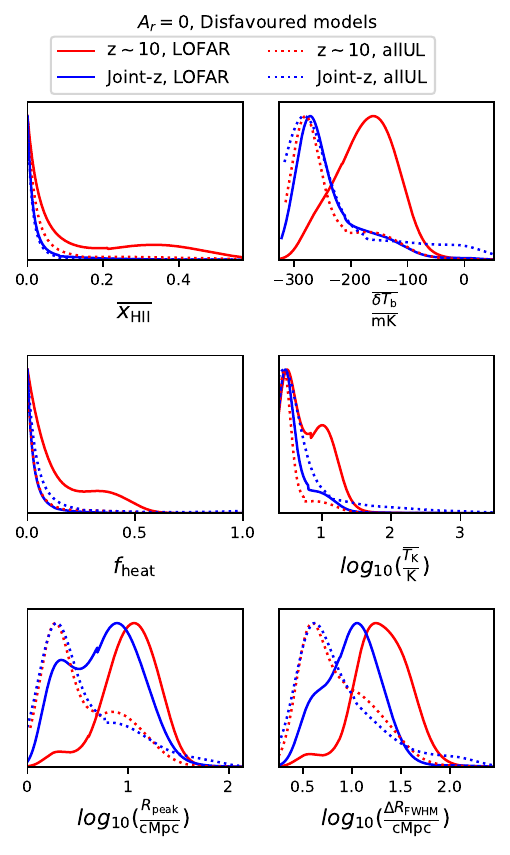}
    \caption{Marginalized probability distribution of the IGM parameters of the disfavoured models at $z\sim10$ when considering upper limits from different radio interferometric observations. These are for the $A_r=0$ scenario. The red and blue curves refer to the {\it Single-z} and {\it Joint-z} cases, respectively. The solid curves represent results when we consider LOFAR upper limits only, while the dotted refers to cases where we have used all available upper limit results from different radio interferometric observations.}
   \label{image_IGM1dall_Ar0z10all}
\end{center}
\end{figure}

The 1D marginalized probability distribution of the disfavoured IGM parameters for the $A_r=0$ and {\it allUL} case at $z\sim 8$ are shown in Fig. \ref{image_IGM1dall_Ar0z8all} while Fig. \ref{image_IGM1dall_Ar0z10all} shows the same but for $z\sim 10$. The disfavoured models at $z\sim8$ in the {\it allUL}, {\it Single-z} and $A_r=0$ scenario have $\AVXHII \lesssim 0.05 ~(0.78)$,  $\AVTK \lesssim 10 ~(2180)$ K, $-352 ~(-375) \lesssim \AVTB \lesssim 3.2 ~(42)$  mK, $ \RPEAK \lesssim 17 ~(169) ~\mpcbyh$ and $ \RFWHM\lesssim 112 ~(316)~\mpcbyh$, while $\FHEAT$ spans the entire range at 68 (95) per cent credible intervals. The disfavoured models at $z\sim 10$ in the {\it allUL} {\it Single-z} case have $\AVXHII \lesssim 0.01 ~(0.35)$,  $\AVTK \lesssim 5 ~(12)$ K, $-285 ~(-285) \lesssim \AVTB \lesssim -241 ~(-117)$  mK, $\FHEAT \lesssim 0.03 (0.36)$ $ \RPEAK \lesssim 7 ~(17) ~\mpcbyh$ and $ \RFWHM\lesssim 12 ~(36)~\mpcbyh$ at 68 (95) per cent credible intervals. As expected, the disfavoured IGM parameter space is much larger in the  {\it allUL} case compared to the LOFAR-only case at $z\sim 8$ and 10, as the MCMC analysis is mainly driven by the strong HERA upper limits. The 68 and 95 per cent credible intervals for the {\it Joint-z} {\it allUL} case are shown in Table \ref{tab_mcmc_igmAr0allUL}. The numbers show that the results for the {\it Joint-z} MCMC analysis are mainly driven by the HERA upper limits at $z\sim8$.

\begin{figure}
\begin{center}
\includegraphics[scale=1.]{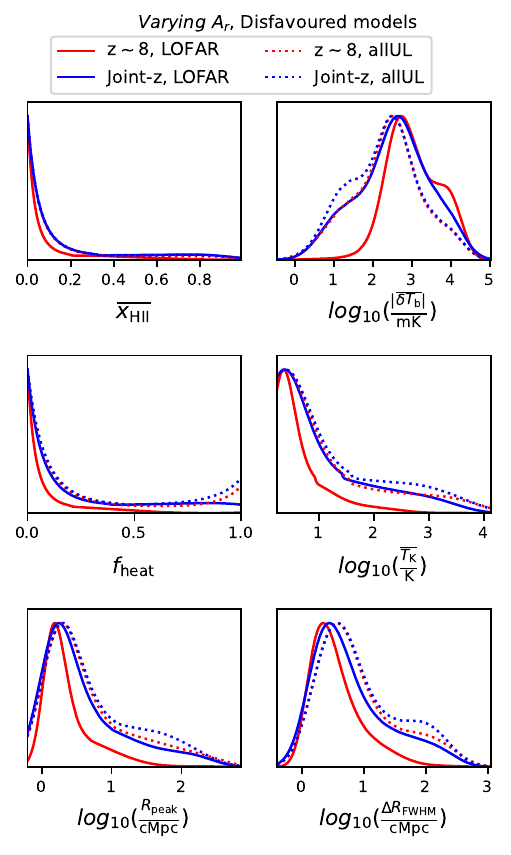}
    \caption{Same as Fig. \ref{image_IGM1dall_Ar0z8all} but for the $Varying~ A_r$ scenario.}
   \label{image_IGM1dall_Arz8all}
\end{center}
\end{figure}

\begin{figure}
\begin{center}
\includegraphics[scale=1.]{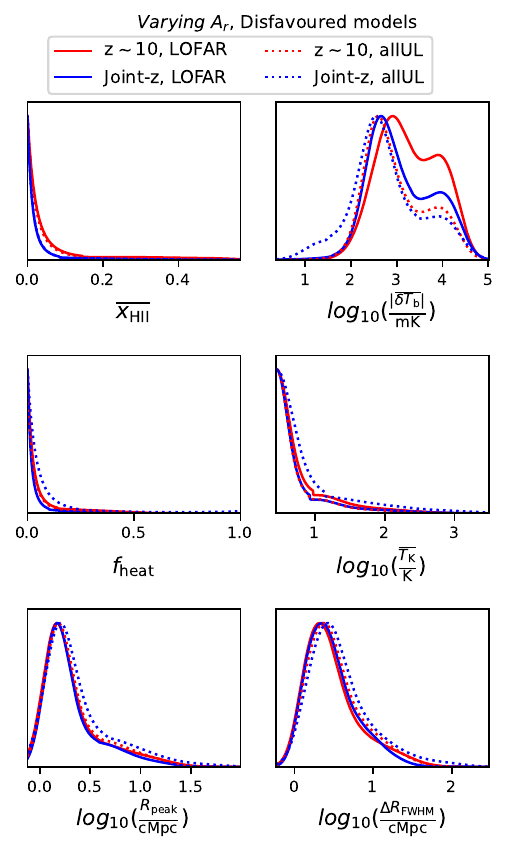}
    \caption{Same as Fig. \ref{image_IGM1dall_Ar0z10all} but for $Varying~ A_r$ scenario.}
   \label{image_IGM1dall_Arz10all}
\end{center}
\end{figure}

Next we consider the {\it Varying $A_r$} scenario, starting with the {\it Single-z} and $\it allUL$ case. In this case, the disfavoured models at $z\sim 8$ show 68 per cent credible intervals on the source parameters as $\zeta\lesssim 1.7$,$\MMIN \gtrsim 1.4\times 10^{10} ~\MSUN$, while the credible intervals of other parameters span the entire range. The same intervals for $z\sim 10$ are $\zeta\lesssim 3.3$,$\MMIN \gtrsim 1.6\times 10^{10} ~\MSUN$ and $\FX \lesssim 5.5$, while the other two parameters' credible intervals span the entire ranges. The 1D marginalized probability distribution of the disfavoured IGM parameters for this case at $z\sim 8$ and 10 is shown in Fig. \ref{image_IGM1dall_Arz8all} and \ref{image_IGM1dall_Arz10all}, respectively. The disfavoured models at $z\sim8$ have $\AVXHII \lesssim 0.05 ~(0.78)$,  $\AVTK \lesssim 10 ~(2180)$ K, $28 ~(2) \lesssim |\AVTB| \lesssim 1905 ~(14791)$  mK, $ \RPEAK \lesssim 8 ~(138) ~\mpcbyh$ and $ \RFWHM\lesssim 16 ~(213)~\mpcbyh$, while $\FHEAT$ spans the entire range at 68 (95) per cent credible intervals for the {\it Single-z} analysis case. The disfavoured models at $z\sim 10$ in the {\it allUL} and {\it Varying $A_r$} case have $\AVXHII \lesssim 0.01 ~(0.32)$,  $\AVTK \lesssim 5 ~(27)$ K, $132 ~(91) \lesssim |\AVTB| \lesssim 9120 ~(24888)$  mK, $\FHEAT \lesssim 0.03 (0.31)$ $ \RPEAK \lesssim 4 ~(11) ~\mpcbyh$ and $ \RFWHM\lesssim 5 ~(20)~\mpcbyh$ at 68 (95) per cent credible intervals for the {\it Single-z} analysis case. 

The 68 and 95 per cent credible intervals for the {\it Joint-z}, {\it allUL} case under the $Varying ~A_r$ scenario are shown in Table \ref{tab_mcmc_igmArallUL}. Similarly to the $A_r=0$ scenario above, we find that the {\it Joint-z} MCMC analysis is also mostly driven by HERA upper limits at $z\sim8$.

\begin{table}
\centering
\caption[]{Constraints on the IGM parameters at three different redshifts for the {\it Varying $A_r$} and {\it allUL} scenario for both the {\it Single-z} and {\it Joint-z} analysis.}  
\small
\tabcolsep 3pt
\renewcommand\arraystretch{2.}
   \begin{tabular}{| c | c | c c c c |}
\hline

\makecell{IGM  \\ Parameters \\(MCMC  \\ Model)}  & \makecell{Credible  \\ intervals}  & $z\sim8$ & $z\sim9$ & $z\sim10$ &   \\

\hline
\multirow{2}{4em}{ $\AVXHII$ ({\it Single-z})}
& 68$\%$ & [0, 0.05] & [0, 0.02] & [0, 0.01]  & \\
\cline{2-6}

& 95$\%$ & [0, 0.78] & [0, 0.46] & [0, 0.32] & \\
\cline{1-6}
\multirow{2}{4em}{$\AVXHII$ ({\it Joint-z})}
&  68$\%$ & [0, 0.06] & [0, 0.02] & [0, 0.004]  & \\
\cline{2-6}

& 95$\%$ & [0, 0.8] & [0, 0.42] & [0, 0.16] & \\

\hline

\multirow{2}{4em}{$\FHEAT$ ({\it Single-z})}

& 68$\%$ & X & [0, 0.06] & [0, 0.03] & \\ 
\cline{2-6}

& 95$\%$ & X & [0, 0.5] & [0, 0.31] & \\
\cline{1-6}
\multirow{2}{4em}{$\FHEAT$ ({\it Joint-z})}
&  68$\%$ & [X] & [0, 0.12] & [0, 0.03] & \\ 
\cline{2-6}

& 95$\%$ & [X] & [X] & [X] & \\
\hline
\multirow{2}{4em}{ $\AVTK$ (K) ({\it Single-z})}

& 68$\%$ & [1.8, 10.0] & [2.2, 5] & [2.6, 5]  & \\ 
\cline{2-6}

& 95$\%$ & [1.8, 2180] & [2.2, 55] & [2.6, 27]  & \\
\cline{1-6}
\multirow{2}{4em}{ $\AVTK$ (K) ({\it Joint-z})}
& 68$\%$ & [1.8, 20] & [2.2, 6] & [2.6, 6]  & \\ 
\cline{2-6}

& 95$\%$ & [1.8, 2089] & [2.2, 371] & [2.6, 110]  & \\

\hline

\multirow{2}{4em}{|$\AVTB$| (mK) ({\it Single-z})}

& 68$\%$ & [28, 11905] & [131,3812] & [132, 9120]  & \\ 
\cline{2-6}

& 95$\%$ & [2, 14791] & [62, 28184] & [91, 24888] & \\
\cline{1-6}
\multirow{2}{4em}{|$\AVTB$| (mK) ({\it Joint-z})}
& 68$\%$ & [18, 1584] & [51,2238] & [77, 2630]  & \\ 
\cline{2-6}

& 95$\%$ & [3, 15848] & [12, 19952] & [19, 31622] & \\

\hline
\multirow{2}{4em}{$\RPEAK$ ($\mpcbyh$) ({\it Single-z})}

& 68$\%$ & [0, 8.0] & [0,5] & [0,4.0]  & \\ 
\cline{2-6}

& 95$\%$ & [0, 138] & [0, 17] & [0, 11] & \\
\cline{1-6}
\multirow{2}{4em}{$\RPEAK$ ($\mpcbyh$) ({\it Joint-z})}
& 68$\%$ & [0, 20] & [0, 5] & [0, 3.0]  & \\ 
\cline{2-6}

& 95$\%$ & [0, 138] & [0, 35] & [0, 15] & \\

\hline
\multirow{2}{4em}{$\RFWHM$ ($\mpcbyh$) ({\it Single-z})}

& 68$\%$ & [0, 16] & [0, 9] & [0,5]  & \\ 
\cline{2-6}

& 95$\%$ & [0, 213] & [0, 76] & [0, 20] & \\
\cline{1-6}
\multirow{2}{4em}{$\RFWHM$ ($\mpcbyh$) ({\it Joint-z})}
& 68$\%$ & [0, 20] & [0, 9] & [0,6]  & \\ 
\cline{2-6}

& 95$\%$ & [0, 218] & [0,76] & [0, 30] & \\

\hline

\end{tabular}
\label{tab_mcmc_igmArallUL}
\end{table}

\section{Inference using \citet{2024MNRAS.534L..30A} upper limit results }
\label{res:appe1}

\begin{table}
\centering
\caption[]{Upper limit on $\DTB$ as reported in \citet{2024MNRAS.534L..30A} at $z=9.1$ for different $k$-bins. The different columns show $k-$bins, mean value $\DTBANSMEAN (k, z)$ and corresponding $1\sigma$ error $ \DTBANSSIGMA (k, z)$. } 
\small
\tabcolsep 8pt
\renewcommand\arraystretch{1.5}
   \begin{tabular}{|c c c|}
\hline
$k$ ($\hbympc$) & $\DTBANSMEAN$ (mK$^2$) & $ \DTBANSSIGMA$ (mK$^2$)  \\
\hline
0.076 & $(5.98)^2$ &  $(15.42)^2$  \\
0.101 & $(8.16)^2$ & $(18.7)^2$  \\
0.133 & $(10.77)^2$ & $(20.89)^2$  \\
0.181 & $(15.24)^2$ & $(25.13)^2$  \\
0.240 & $(17.6)^2$ & $(24.47)^2$  \\
0.319 & $(20.88)^2$ & $(28.15)^2$ \\
0.436  & $(27.12)^2$ & $(41.63)^2$  \\

\hline
\end{tabular}
\label{tab_obs_ansh}
\end{table}

\begin{figure}
\begin{center}
\includegraphics[scale=0.4]{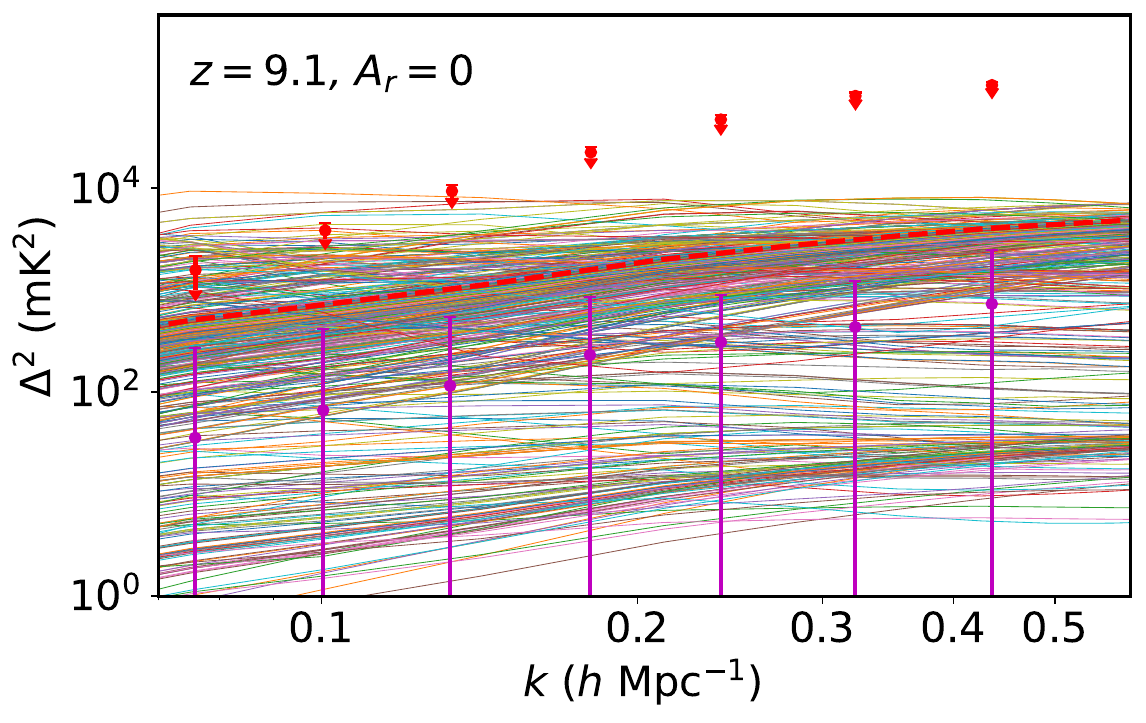}
    \caption{A subset of 1000 power spectra simulated with {\sc grizzly} at $z=9.1$.
    These power spectra correspond to those with $A_r = 0$. 
    The red and magenta points refer to the recent LOFAR 1$\sigma$ upper limits from \citet{lofar2024} and \citet{2024MNRAS.534L..30A}, respectively. As a reference, the red dashed line corresponds to the power spectrum for a completely neutral IGM with no X-ray heating and $\TS=\TK$.}
   \label{image_psts_z_ansh}
\end{center}
\end{figure}

\begin{figure}
\begin{center}
\includegraphics[scale=0.8]{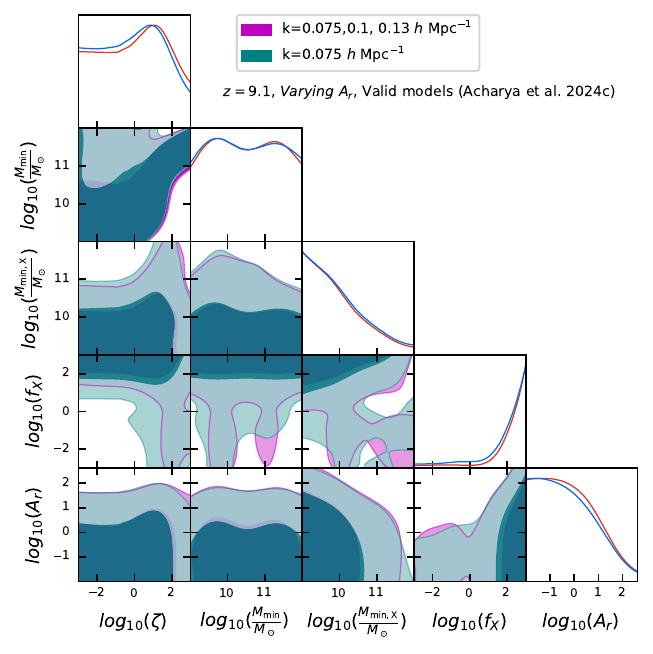}
    \caption{Posterior distribution of the sources parameters of the models which are favoured by the LOFAR upper limits from \citet{2024MNRAS.534L..30A} at $z=9.1$ for the {\it Varying $A_r$} scenario. Magenta indicates the case in which upper limits in three $k-$bins are considered, while teal represents the case when only one $k-$bin is used. The contour levels in the two-dimensional contour plots refer to the $1\sigma$ and $2\sigma$ credible intervals of the models favoured by the LOFAR upper limit at this redshift. The diagonal panels represent the corresponding marginalized probability distributions of each parameter. }
   \label{image_sourceparam_Ar_ansh}
\end{center}
\end{figure}

The LOFAR upper limits considered in the main text, as well as those obtained by \citet{2020MNRAS.493.1662M}, contain excess noise, and thus are conservative limits in nature. Recently, \citet{2024MNRAS.534L..30A} have re-analysed   the 10 nights of data used in \cite{2020MNRAS.493.1662M} with a different approach, and reported a much stronger 2-$\sigma$ upper limit of $\DTB(k = 0.075 \hbympc, ~z=9.1, \rm 141 ~hours) \approx (25)^2 ~{\rm mK}^2$ (see Table \ref{tab_obs_ansh}). This was obtained by assuming an accurate bias correction for the excess noise by subtracting the excess component. As this approach increases the risk of signal loss, more testing is required to verify the robustness of the results. Nevertheless, here we use these upper limits to explore their impact on the derived constraints for the physical state of the IGM.

Figure \ref{image_psts_z_ansh} shows 1000 randomly chosen models from the {\sc grizzly} simulation database, together with the LOFAR 1$\sigma$ upper limits from \citet[][red]{lofar2024} and \citet[][magenta, $\DTBANSMEAN$ with error $\DTBANSSIGMA$]{2024MNRAS.534L..30A}.  We see that the red dashed curve, corresponding to the power spectrum of a completely neutral and unheated IGM at $z=9.1$ obtained assuming constant values of  $x_{\rm HI}=1$ and $\TS=\TK=2$~K, remains above the 1$\sigma$ upper limits from \citet{2024MNRAS.534L..30A} at least for one $k-$bin. Unlike the results presented in the main text, here we find that all models are above $\DTBANSMEAN-\DTBANSSIGMA$. 

When performing the same MCMC analysis discussed in the main text to identify the disfavoured models, we find that the marginalized probabilities are high over the full parameter space, indicating that a large number of models are disfavoured. For this reason, differently from the analysis performed in the main text, here we choose to study the properties of the favoured models by adopting for the likelihood of a model surviving at $z$ the expression $\mathcal{L}_{\mathrm{su, single}-z}(\bm{\theta}, z)=1-\mathcal{L}_{\mathrm{ex, single}-z}(\bm{\theta}, z)$.

We perform the {\it Single-z} analysis following two approaches, the first of which is the same one adopted in the main text, i.e. we consider the upper limits at the three smallest $k-$bins. In the second case, instead, we consider only the smallest $k-$bin upper limit, which is the strongest and thus it is the one which mainly affects the constraints on the source and IGM parameters. Using only one $k-$bin also have the advantage of breaking the correlation originating from the approach adopted in \citet{2024MNRAS.534L..30A} when using upper limits at different $k-$bins.

Figure \ref{image_sourceparam_Ar_ansh} shows the posterior distribution of the source parameters for the {\it Varying $A_r$} scenario at $z= 9.1$ obtained from the MCMC analysis using the upper limits from \citet{2024MNRAS.534L..30A} on the allowed models. The 68 per cent credible intervals limits of the allowed models are  $\zeta \lesssim 14$, $\MMIN \lesssim 5\times 10^{11} ~\MSUN$, $\MMINX \lesssim 1.5\times 10^{10} ~\MSUN$, $f_X \gtrsim 100$ and $A_r \lesssim 2$ for the {\it {\it Varying} $A_r$ scenario}. The limits are similar when we use one upper limit at $k=0.075 \hbympc$.

\begin{figure}
\begin{center}
\includegraphics[scale=0.7]{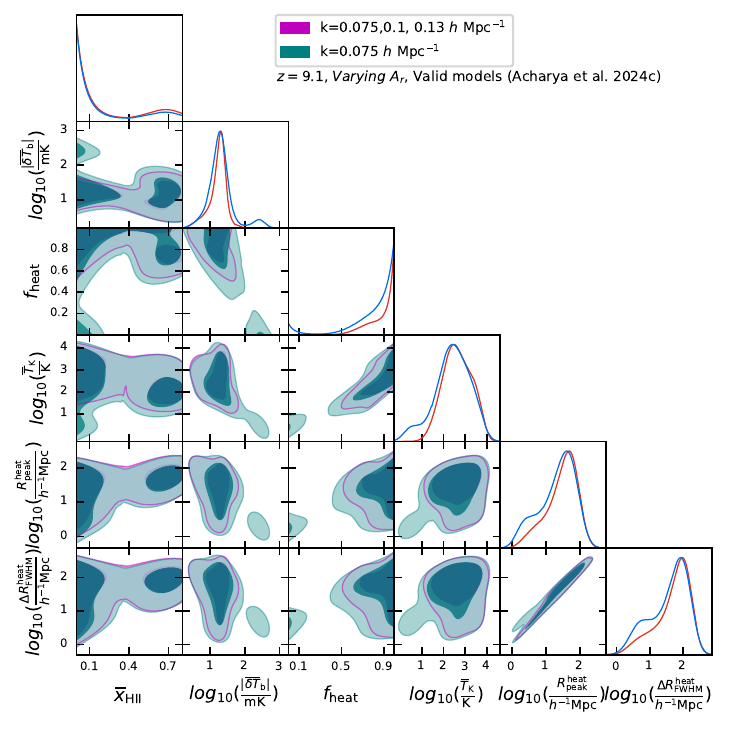}
\caption{Posterior distribution of the IGM parameters of the models which are favoured by the LOFAR upper limits from 
\citet{2024MNRAS.534L..30A} at $z=9.1$ for the {\it Varying $A_r$} scenario. Magenta indicates the case in which upper limits in three $k-$bins  are considered, while teal represents the case when only one $k-$bin is used. The contour levels in the two-dimensional contour plots refer to the 1$\sigma$ and 2$\sigma$ credible intervals of the models favoured by the LOFAR upper limits at this redshift. The diagonal panels represent the corresponding marginalized probability distributions of each parameter. }
   \label{image_igmparam_Ar_ansh}
\end{center}
\end{figure}

Figure \ref{image_igmparam_Ar_ansh} shows the posterior distribution of the IGM parameters of the favoured models obtained from the {\it Single-z} analysis with the upper limits from \citet{2024MNRAS.534L..30A}  at $z=9.1$. The 95 per cent credible intervals of the allowed models' IGM parameters are $\AVXHII \lesssim 0.75$,  $27 \lesssim \AVTK \lesssim 7\times10^3$ K, $\FHEAT \gtrsim 0.62$, $4 \lesssim \AVTB \lesssim 44$ mK, $ 3 \lesssim \RPEAK\lesssim 160 ~\mpcbyh$ and $ 4 \lesssim \RFWHM\lesssim 316~\mpcbyh$. The limits on the IGM parameters for the case where we used the upper limit at $k=0.075 \hbympc$ only are similar to the case where we used three $k-$bins upper limits. We note that, similarly to the scenarios considered in the main text, these constraints are also highly dependent on the chosen priors on the source parameters (see Appendix \ref{res:appe2}).

\end{appendix}

\end{document}